\documentstyle[epsfig]{mn2e}

\title[Void Finder Comparison Project]{The Aspen--Amsterdam Void Finder Comparison Project}
\author[J.\ M.\ Colberg et al.]
       {J\"org M.\ Colberg$^{1,2}$, Frazer Pearce$^3$, Caroline Foster$^{4,5}$,
        Erwin Platen$^{6}$, \newauthor
        Riccardo Brunino$^3$, Mark Neyrinck$^{7}$, Spyros Basilakos$^{8}$, Anthony
	      Fairall$^{9}$, \newauthor Hume Feldman$^{10}$,Stefan Gottl\"ober$^{11}$,
        Oliver Hahn$^{12}$, Fiona Hoyle$^{13}$, \newauthor
	      Volker M\"uller$^{11}$,	Lorne Nelson$^4$, Manolis Plionis$^{14,15}$, Cristiano
	      Porciani$^{12}$, \newauthor
	      Sergei Shandarin$^{10}$, Michael S. Vogeley$^{16}$, Rien van de Weygaert$^{6}$\\
	      $^1$ Carnegie--Mellon University, 5000 Forbes Ave, Pittsburgh, PA 15213, USA\\
        $^2$ Astronomy Department, University of Massachusetts, Amherst, 
             MA 01003, USA\\
        $^3$ School of Physics \& Astronomy, University of 
             Nottingham, Nottingham, NG7 2LE, UK\\
        $^4$ Bishop's University, Department of Physics, 2600 College Street,
             Sherbrooke, QC J1M 0C8, Canada\\
        $^{5}$ Centre for Astrophysics \& Supercomputing, Swinburne
	             University of Technology, Hawthorn VIC 3122, Australia\\
        $^{6}$ Kapteyn Institute, University of Groningen, PO Box 800, 
             9700 AV Groningen, The Netherlands\\
        $^{7}$ Institute for Astronomy, University of Hawaii, Honolulu, HI 96822, USA\\
        $^8$ Research Center for Astronomy \& Applied Mathematics, Academy of Athens, 
             Soranou Efessiou 4, 11-527, Athens, Greece\\
        $^9$ Department of Astronomy, University of Cape Town, Private Bag,
              Rondebosch 7700, South Africa\\
        $^{10}$ Department of Physics and Astronomy, University of Kansas, 
             Lawrence, KS 66045, USA\\
        $^{11}$ Astrophysikalisches Institut Potsdam, An der Sternwarte 16,
             14482 Potsdam, Germany\\
        $^{12}$ ETH Z\"urich, 8093 Z\"urich, Switzerland\\
        $^{13}$ Department of Physics and Astronomy, Widener University, One
                University Place, Chester, PA 19013\\
        $^{14}$ Institute of Astronomy \& Astrophysics, National Observatory of Athens, 
              I.\ Metaxa \& B.\ Pavlou, P.\ Penteli 152 36, Athens, Greece\\
        $^{15}$ Instituto Nacional de Astrofisica, Optica y
                Electronica (INAOE) Apartado Postal 51 y 216,
                72000 Puebla, Pue., Mexico\\
        $^{16}$ Department of Physics, Drexel University, 3141 Chestnut Street, 
             Philadelphia, PA 19104, USA\\
}
\date{Accepted 200? ???? ??.
      Received 2007 ???? ??;
      in original form 2007  xx}

\pagerange{\pageref{firstpage}--\pageref{lastpage}}
\pubyear{200?}

\begin{document}

\maketitle

\label{firstpage}

\begin{abstract}

Despite a history that dates back at least a quarter of a century
studies of voids in the large--scale structure of the Universe are
bedevilled by a major problem: there exist a large number of quite
different void--finding algorithms, a fact that has so far got in the
way of groups comparing their results without worrying about whether
such a comparison in fact makes sense. Because of the recent increased
interest in voids, both in very large galaxy surveys and in detailed
simulations of cosmic structure formation, this situation is very
unfortunate. We here present the first systematic comparison study of
thirteen different void finders constructed using particles, haloes,
and semi--analytical model galaxies extracted from a subvolume of the
Millennium simulation. The study includes many groups that have
studied voids over the past decade. We show their results and discuss
their differences and agreements. As it turns out, the basic results
of the various methods agree very well with each other in that they
all locate a major void near the centre of our volume.  Voids have very
underdense centres, reaching below 10 percent of the mean cosmic
density. In addition, those void finders that allow for void galaxies show
that those galaxies follow similar trends. For example, the
overdensity of void galaxies brighter than $m_B = -20 $ is found to be
smaller than about $-0.8$ by all our void finding algorithms.

\end{abstract}

\begin{keywords}
cosmology: theory, methods: N-body simulations, dark matter, large-scale structure of Universe
\end{keywords}

\section{Introduction}

Large regions of space that are only sparsely populated with galaxies,
so--called voids, have been known as a feature of galaxy surveys since
the first of those surveys was compiled, the most well--known cases 
being the famous void in Bo\"otes, discovered by Kirshner et al. (1981),
and the first void sample of de Lapparent, Geller \& Huchra (1986).
However, due to the fact that voids occupy a large fraction of space,
only recently have galaxy surveys become large enough to allow systematic
studies of voids and the galaxies inside them. For
recent studies of voids and void galaxies in the two--degree field Galaxy
Survey (2dFGRS, Colless et al. 2001) and the Sloan Digital Sky Survey (SDSS,
York et al. 2000) see Hoyle \& Vogeley (2004), Ceccarelli et al. (2006),
Patiri et al. (2006a), Tikhonov (2006),
von Benda-Beckmann \& M\"uller (2008) and Rojas et al. (2004), 
Goldberg et al. (2005), Hoyle et al. (2005), Rojas et al. (2005), 
Patiri et al. (2006b), 
respectively. Also see Croton et al. (2004) 
for a recent, detailed study of the void probability function in the 2dF.

On the theoretical side, progress has been mirrored by vast improvements 
in models and simulations, with systematic studies of large numbers of
voids now being common (see, for example, the early works by Regos \&
Geller 1991, Dubinski et al. 1993 or Van de Weygaert \& Van Kampen
1993, and the more recent Arbabi-Bidgoli \& M\"uller 2002, Mathis \& 
White 2002, Benson et al. 2003, Gottl\"ober et al. 2003, Goldberg \& 
Vogeley 2004, Sheth \& Van de Weygaert 2004, Bolejko et al. 2005, 
Colberg et al. 2005, Padilla et al. 2005, Brunino et al. 2006, 
Furlanetto \& Piran 2006, Hoeft et al. 2006, Lee \& Park 2006, 
Park \& Lee 2006, Patiri et al. 2006c, Shandarin et al. 2006). Theory 
shows that voids are a real feature of large--scale structure, since 
initially underdense regions grow in size as overdense regions collapse
under their own gravity (see, for example, Sheth \& Van de Weygaert 2004).
But while the general picture appears to be well supported by the
standard $\Lambda$ Cold Dark Matter ($\Lambda$CDM) cosmology, Peebles 
(2001) pointed out some potentially critical issues. Does the $\Lambda$CDM 
cosmology produce too many objects in voids that have no observational
counterparts? A detailed discussion of other reasons why studies of voids 
are an interesting topic is beyond the scope of this paper. Briefly, 
their role as a prominent feature of the Megaparsec Universe means that 
a proper and full understanding of the formation and dynamics of the 
Cosmic Web is not possible without probing the structure and evolution 
of voids. A second rationale is that of inferring global cosmological 
information from the structure and geometry of and outflow from voids. 
The third aspect is that of providing a unique and still largely pristine
environment for testing theories of the formation and evolution of galaxies.

Despite the growing interest in voids and the large number of recent
studies, a fairly significant problem remains: as it turns out, almost
every study uses its own void finder. There is general agreement that
there are voids in the data or in the simulations, but many different
ways were proposed to find them. Thus, the resulting voids are either
spherical (with or without overlap), shaped like lumpy
potatoes\footnote{JMC admits that this picture, while being accurate,
is not very pretty.}, or they percolate all across the studied volume.
What is more, some groups do not allow for the existence of void
galaxies, whereas many others do. An added complication is that many
theoretical studies use the dark matter distribution to find voids,
whereas observational studies have to rely on galaxies. As a
consequence, it is not clear how the results from studies done by
different groups can be compared, especially if observational and
theoretical results are brought together. What most studies so far can
agree on is that a) voids are very underdense in their centres
(approaching around five percent of the mean density) and that b)
voids often have very steep edges.  In other words, the number of both
observed and simulated galaxies increases very rapidly when reaching
the edge of a void, and the corresponding result has been found for
the density of dark matter in studies that used dark--matter only
simulations.

Given the disagreements in the different methods, which are in part due to
the different nature of the data sets used, the aim of this work is 
very modest. We apply thirteen different void finders, all of which have
been used over the past decade to study voids, to the same data set 
in order to compare the results. As our data set we use particles, haloes, and
semi--analytical model galaxies (Croton et al. 2005) from a subvolume of 
the Millennium simulation (Springel et al. 2005) specifically selected
to be underdense and therefore void-rich. That way, while the
methods are as different as finding connected cells on a density grid and
identifying empty regions in the model galaxy distribution by eye, a
meaningful comparison is still possible, since each void finder treats
a subset of the same data set. 

The aim of this paper is {\it not} to argue which void finder provides the
best way to identify voids. We do hope, however, that this paper will
allow the reader to understand the differences between the different
void finders so that it will be easier to compare different studies of
voids in the literature. We also hope that this paper will trigger 
more detailed follow--up studies to work towards a more unified view of
this topic and to study properties of voids not covered here, such as
for example their shapes and orientations, in detail.

This paper is organised as follows: in Section 2 we detail the
simulation from which the test region was extracted and describe the
procedure each group was asked to undertake. In Section 3 we briefly
describe each void finding algorithm, before we undertake a comparison
of the voids found in Section 4. Section 5 contains a summary and
conclusions.

\section{The Simulation and extraction procedure} \label{simulation}

For this work, we use the Millennium simulation (Springel et al. 2005) 
and a matched $z=0$ galaxy catalogue, created using a semi--analytical galaxy
formation model (Croton et al. 2005). The simulation of the 
concordance $\Lambda$CDM cosmology contains 2160$^3$ particles
in a (periodic) box of size 500\,$h^{-1}$\,Mpc in each dimension.
The cosmological parameters are total matter density $\Omega_m = 0.25$, 
dark energy/cosmological constant $\Omega_\Lambda=0.75$, Hubble constant 
$h = 0.73$, and the normalisation of the power spectrum $\sigma_8 = 0.9$. 
With these parameters, each dark matter particle has a mass of 
$8.6 \times 10^8\,h^{-1}$\,M$_{\odot}$. 

In the Millennium simulation volume we located a 60\,$h^{-1}$\,Mpc region
centred on a large void and extracted the coordinates of the 12,528,667 
dark matter particles contained within it. This subvolume thus has a 
mean density which is lower than the cosmic mean, corresponding to an 
overdensity $\delta = \rho/\bar{\rho} - 1 = -0.28$.

We also extracted a list of the 17,604 galaxies together with their 
BVRIK dust corrected magnitudes (down to B=-10) that are present in the semianalytic 
catalogue of Croton et al. (2005) within this volume and the 4,006 dark 
matter halos present in the subfind catalogue (a clean spherical overdensity 
based catalogue) with masses greater than $10^{11}\,h^{-1}$\,M$_{\odot}$. 
Note that while the small volume prohibits statistical comparisons
between void finders it allows for void--by--void comparisons.

Each group was asked to run their void finder with their preferred parameters 
on this database and return a void list for the voids found, tagging each of 
the dark matter particles, galaxies, and haloes with the void identifier of 
the void they resided in. This allowed simple plotting and analysis of each 
void sample. For overlapping voids, the dark matter particle was to be 
assigned to the larger void. As the region is not periodic we only requested 
information about voids whose centres lay within the central 
40\,$h^{-1}$\,Mpc region (i.e. the outer 10\,$h^{-1}$\,Mpc was to be neglected). 

The top left and top centre panels of Figure~\ref{fig:void_gals1} show 
slices of thickness 5\,$h^{-1}$\,Mpc through this central region. The top 
left panel only contains the distribution of the dark 
matter, whereas the top centre panel includes model galaxies on top of the 
dark matter distribution. The largest halo has a mass of only $1.75\times 
10^{12}\,h^{-1}$\,M$_\odot$, so filaments in these images correspond only 
to the less massive filaments in standard slices through the dark matter 
distribution as seen in, for example, Springel et al. (2005). Furthermore, 
the slice contains a total of 145,194 dark matter particles, equivalent to 
an overdensity of $\delta = -0.77$. It is important to keep these numbers 
in mind when studying the results obtained by the various void finders.
Subsequent panels show the largest void identified by each group 
and those galaxies contained within all voids identified.

\begin{table*}
\begin{tabular}{|lll|}
\hline
{\bf Author} & {\bf Base} & {\bf Method} \\ 
\hline
Brunino          & Haloes                    & Spherical regions in halo distribution \\
Colberg          & Dark matter density field & Irregularly shaped underdense regions around local density minima \\
Fairall          & Galaxies                  & Empty regions in galaxy distribution \\
Foster/Nelson    & Galaxies                  & Empty regions in galaxy distribution \\
Gottl\"ober      & Haloes/Galaxies           & Spherical empty regions in point set \\
Hahn/Porciani    & Dark matter density field & Tidal instability in smoothed density field \\
Hoyle/Vogeley    & Galaxies                  & Empty regions in galaxy distribution \\
M\"uller         & Halos/Galaxies            & Empty convex regions in point set \\ 
Neyrinck         & Dark matter density field & ZOBOV, depressions in unsmoothed DM field \\
Pearce           & Dark matter               & Local density minima spheres \\
Platen/Weygaert  & Dark matter density field & Watershed DTFE \\
Plionis/Basilakos & Dark matter density field & Connected underdense density grid cells\\
Shandarin/Feldman& Dark matter density field & Connected underdense density grid cells\\
\hline
\end{tabular}
\caption{An overview of the void finders used in this study.}
\label{tab:voidfinders}
\end{table*}

\section{Void Finders} \label{voidfinders}

This section gives a brief outline of the void finders used for this
study, grouped into those which rely on the dark matter distribution
and those which rely on the sparser galaxy or halo distributions (also
see Table~\ref{tab:voidfinders} for a general overview). For
more details, the interested reader is referred to the individual 
studies by the different groups. Anyone simply interested in the 
results can skip to the next section. Please note that in this study,
all group finders use real--space data.

\subsection{Finders based on the dark matter distribution}

\subsubsection{Colberg: Irregularly shaped underdense regions around local
               density minima}

This method was introduced in Colberg et al. (2005), where it was used 
to study voids in the dark matter distribution of a suite of large N--body 
simulations. The starting point for Colberg et al.'s void finder is the 
adaptively smoothed distribution of the full dark matter distribution in 
a simulation. Proto--voids are constructed in a fashion quite similar to 
Hoyle \& Vogeley's void finder, the difference being that Colberg's 
uses local minima in the density field as the centres of voids, and the 
mean density of the spherical proto--voids is required to be smaller or 
equal to an input threshold, which, following a simple linear theory
argument, is taken to be $\delta = -0.8$ (Blumenthal et al. 1992). 
Proto--voids are then merged according to a set of criteria, which 
allow for the construction of voids that can have any shape, as long as 
two large regions are not connected by a thin tunnel (which would make 
the final void look like a dumbell). The voids thus can have arbitrary 
shapes, but they typically look like lumpy potatoes.

For this study, a grid of size $480^3$ and a minimum void radius of
$r_{\mbox{{\footnotesize min}}} = 2.0\,h^{-1}$\,Mpc were used. In the
following, this void finder and its results are referred to as {\it Colberg}.

\subsubsection{Pearce: Spheres around local density minima}

For every particle in the Millennium simulation local densities 
were calculated by smoothing over the nearest 32 neighbours using a 
beta spline kernel (Monaghan \& Lattanzio 1985). This list was then 
ranked in density order (starting from the most underdense particle), 
and independent initial void centres were chosen such that they were 
more than 2\,$h^{-1}$\,Mpc away from a 
previously selected centre (up to $\delta = -0.965$). The radial 
distribution of particles about these trial centres was then used to 
calculate the first up--crossing above $\delta = -0.9$.
These radii were then sorted in size order, and the resulting list was 
cleaned by removing voids whose centre lay within an already found void. 
The 3,024 voids found by this procedure were used as the starting points 
for the more traditional halo based group finder used by Brunino et al. 
(2007). In the following, this void finder and its results are referred 
to as {\it Pearce}.

\subsubsection{Hahn/Porciani: Equation of motion in smoothed density field}

A stability criterion for test--particle orbits is used to discriminate
four environments with different dynamics (Hahn et al. 2006). The
classification scheme is based on a series expansion of the equation
of motion for a test particle in the smoothed matter distribution. The
series expansion yields a zero order term, the acceleration, 
and a second order term, the tidal field $T_{ij}$ (Hessian of
the potential). The eigenvalues of $T_{ij}$ characterise the triaxial
deformation of an infinitesimal sphere due to the gravitational
forces. Voids are classified as those regions of space where $T_{ij}$ has
no positive eigenvalues (tidally unstable).

The method has one free parameter, the size of the Gaussian filter 
used to smooth the potential. This parameter is set to 
$R_s = 2.09\,h^{-1}$Mpc, which corresponds to a mass of 
$10^{13}\,h^{-1}M_\odot$ contained in the filter at mean density. This
choice gives excellent agreement with a visual classification (see the 
discussion in Hahn et al. 2006). This mass scale corresponds to about 
$2\,M_*$ at $z=0$. 

The tidal field eigenvalues are evaluated on a grid. Then, contiguous
regions classified as voids are linked together. Voids can thus have 
arbitrary shapes, and their volumes are proportional to the number
of cells linked together. In the following, this void finder and its 
results are referred to as {\it Hahn/Porciani}.

\subsubsection{Neyrinck: Zobov}

ZOBOV (ZOnes Bordering On Voidness, Neyrinck (2008)) is an inversion of a
 publicly available halo-finder, VOBOZ\footnote{Available at
 {\it http://ifa.hawaii.edu/$\sim$neyrinck/VOBOZ}.} (Neyrinck et
 al. 2005). 
ZOBOV differs from VOBOZ in that ZOBOV looks for density
 minima instead of maxima, and does not consider gravitational binding.
 ZOBOV has some unique features: it is entirely parameter-free, working
 directly on the unsmoothed particle distribution; and, it returns all
 (even possibly spurious) depressions surrounding density minima, along
 with estimates of the probability that each arises from Poisson noise.

 The first step is density estimation and neighbour identification for
each dark-matter particle, using what Schaap (2007) calls the Voronoi
 Tesselation Field Estimator.  ZOBOV then partitions the particles into
 zones (depressions) around each minimum.  Each particle jumps to its
 lowest-density neighbour, repeating until it reaches a minimum. A
 minimum's {\it zone} is the set of particles which flow downward into
 it.  Zones resemble voids, but because of unsmoothed discreteness
 noise, many zones are spurious, and others are only cores of voids
 detected by eye.  So, ZOBOV must join some zones together to form
 voids.  Voids around each zone grow by analogy with a flooding
 landscape (representing the density field): water flows into
 neighbouring zones, adding them to the original zone's void.  The
 zone's void stops growing when the water spills into a zone deeper
 than the original zone, or the whole field is submerged.  The
 probability that a void is real is judged by the ratio of the density
 at which this happens to the void's minimum density.

 This density contrast $r$ is converted to a probability through
 comparison with a Poisson point distribution; see Neyrinck et al. (2005) for
 details.  The ZOBOV catalogue used for comparison with other
 void-finders includes only voids exceeding a 5-$\sigma$ probability
 threshold, which corresponds to a density contrast of 2.89.  Also,
subvoids exceeding this threshold have been removed from parent voids.
In the following, this void finder and its results are referred to as 
{\it Neyrinck}.

\subsubsection{Platen/Weygaert: Watershed void finder}

The Watershed Void Finder (WVF) is an implementation of the Watershed Transform (WST)
for image segmentation towards the analysis of the Cosmic Web. The Watershed Transform
is a familiar concept in mathematical morphology and was first introduced by
Beucher \& Lantuejoul (1979, also see Beucher \& Meyer 1993).

The WST delineates the boundaries of separate domains, ie. the {\it basins}, into which
which the yields of e.g. rainfall will collect. The analogy with the cosmological context
is straightforward: {\it voids} are to be identified with the {\it basins}, while the
{\it filaments} and {\it walls} of the cosmic web are the ridges separating the voids
from each other.
				    
The voids are computed by an algorithm that mimics the flooding process. First the
cosmological point distribution is transformed by the DTFE technique into a density
field. DTFE (Schaap \& van de Weygaert 2000, Schaap 2007) assures an optimal rendering
of the hierarchical, anisotropic and voidlike nature and aspects of the weblike cosmic
matter distribution. The density field is adaptively smoothed by {\it nearest neighbour
median filtering} (Platen et al. 2007). Minima are selected from the smoothed field and
marked as the sources of flooding. While the ``watershed" level rises a growing fraction
of the ``landscape" will be flood: the basins expand. Ultimately basins will meet at the
ridges, saddlepoints in the density field. These ridges define their boundaries,
and are marked as edge separating the two basins. The procedure is continued until the
density field is completely immersed, leaving a division of the landscape into
individual segments separated by {\it edges}. The edges delineate the skeleton of the field and
outline the {\it voids} in the density field.

The voids in the watershed procedure have no shape constraints. By definition the voids
fill space completely. Nearly without exception galaxies and dark halos are located on the
ridges of the cosmic web, implying a minimal amount of galaxies to be located in the
watershed void segments.

\subsubsection{Plionis/Basilakos: Connected underdense density grid cells}

This void finder is applied on a regular 3-D grid of the DM particle 
distribution or of a smoothed galaxy distribution, and it is
based in identifying those grid cells (which we call ``void cells'') whose 
density contrast lies below a specific threshold. Then all neighbouring 
(touching) ``void cells'' are connected to form candidate voids (see also 
Plionis \& Basilakos 2002). Therefore, by construction, voids do not 
overlap, and they can have an arbitrary shape, which is approximated 
by an ellipsoidal configuration (see Plionis \& Basilakos 2002). 
Of course, increasing the threshold 
one tends to percolate through the available volume by connecting voids.
The threshold below which the ``void cells'' are identified is chosen so 
that a specific fraction of the probability density function (pdf) is 
used. For example, the voids presented here are based on the lowest 
12.5\% density ``void cells'', which corresponds to $\delta\rho/\rho\simeq -0.92$.

In order to identify significant voids from our candidate list we 
compare with voids found in 1000 realizations of the DM particle 
distribution, using again the lowest 12.5\% density void cells of 
each ``random--realization'' pdf. Now a probability curve as a 
function of void size can be built. Smaller voids appear with a large 
frequency in the random realizations and thus a candidate void is 
considered as significant only if its probability of appearing in a 
random distribution is $<0.05$.

There are two free parameters in this void identification procedure: 
(a) The grid cell size and (b) the threshold below which void cells are
identified. The first is selected arbitrarily in this work such that it
roughly encloses the volume of a typical cluster of galaxies, $(2{\rm Mpc})^{3}$,
while the second is selected such that it maximises the number of significant
voids. In the following, this void finder and its results are referred to as 
{\it Plionis/Basilakos}.

\subsubsection{Shandarin/Feldman: Connected underdense density grid cells}

Voids are defined as the individual 3D regions of the low-density excursion set 
fully enclosed with the  isodensity  surfaces (for more details see Shandarin et al. 2006). 
Here, we first generate the density field on a uniform rectangular grid using the cloud-in-cell (CIC) technique. The grid parameter is chosen to be equal to the mean separation of particles in the whole simulation $d=(500/2160) h^{-1}$ Mpc. The CIC algorithm
uses particles of the same size. Then the density field is smoothed with a spherical 
Gaussian filter with 
$R_G=1 h^{-1} $ Mpc, assuming nonperiodic boundary conditions and empty space 
beyond the boundaries. In the analysis we use only the central part of the cube, 
slicing $4.5d$ from every face of the initial cube affected by smoothing. 
The final cube consists of 250$^3$ grid sites with the volume of about 91\% 
of the initial cube. Nonpercolating voids reach  maximum sizes at the percolation transition (Shandarin et al. 2004). The technique makes no assumptions about the 
shapes of voids that generally are highly nonspherical.
At higher thresholds the total volume in all but the percolating void drops off 
precipitously  and the  excursion set practically becomes a single percolating void.
Our  voids are identified at the percolation threshold 
$\delta \approx -0.88$ (filling fraction of the voids, FF$_V$ = 20\%).  
We find  19 voids larger than $ 5 h^{-3}$ Mpc$^3$. The largest void is of irregular
shape and its volume is $V=2.1 \times 10^4 h^{-3}$ Mpc$^3$.
There are neither halos nor galaxies inside these voids. Galaxies start  to appear in the percolating void at $\delta > -0.86$ (FF$_V >  $27\% )
and halos at $\delta > -0.63$   (FF$_V  > $66\%).
In the 
following, this void finder and its results are referred to as 
{\it Shandarin/Feldman}.

\subsection{Finders based on the galaxy or halo distribution}

\subsubsection{Brunino: Spherical voids in halo catalogue}

This void finder algorithm uses the void centres provided by
{\it Pearce}'s algorithm as an initial guess for the location of the underdense
regions. These positions are then used to search for the maximum spheres 
that are empty of haloes with masses larger than 8.6$\times$10$^{11} 
h^{-1}$M$_{\sun}$, or 1000 particles. To characterise the final position 
of the void centres and their radii we populate a sphere of radius $R=5 
h^{-1}$ Mpc,centred on each initial position, with 2000 random points 
(the choice of these quantities has proved to be the most convenient in 
order to obtain a stable result). For every point in this sphere, the 
position of the closest four haloes lying in geometrically ``independent'' 
octants is found. The sphere defined by these four haloes is then built.
This is repeated for all the 2000 random points. As a characterisation 
(position and radius) of the void, the biggest empty spherical region 
generated in the previous step is chosen.

It is important to stress that the position of the void defined in this
way normally does not match the position of the initial guess.
Furthermore, in the present work, voids whose centre turned out to lie
inside a larger void have been discarded. A total of six void regions
have been found in the volume of interest, three of which have been neglected
applying this criteria. This algorithm is a variant of the one described
in Patiri et al. (2006a) which has been developed to resemble the 
observational technique used to detect voids to enable a more direct 
comparison with simulations (e.g. Trujillo et al. 2006, Brunino et al. 2007)
In the following, this void finder and its results are referred to as 
{\it Brunino}.

\subsubsection{Fairall: Voids in the galaxy distribution}

Voids have been located manually by inspection of slice 
visualisations: A moving slice, in $x$ and $y$ with thickness 
$\Delta z = 5\,h^{-1}$\,Mpc, has been passed through the data 
in steps of $2.5\,h^{-1}$\,Mpc. Its progress has been  
visualised by software that shows both individual galaxies 
and large--scale structures, the latter based on minimal 
spanning trees with percolations of $1\,h^{-1}$\,Mpc or less 
(effectively ``friends of friends''). The voids are conspicuous 
cavities, approximately spherical, empty or almost empty of 
galaxies, visible in consecutive slices, with sharply defined 
walls formed by large--scale structures. Since the voids interconnect 
with one another, the large--scale structures do not necessarily 
completely enclose each void. If a void departs from sphericity,
an average radius is estimated. Where the data allow, distinct 
voids as small as $2.5\,h^{-1}$\,Mpc (radius) are identified. In the 
following, this void finder and its results are referred to as 
{\it Fairall}.

\subsubsection{Foster/Nelson: Voids in the galaxy distribution} 

The identification of voids is calculated using a prescription similar 
to that of Hoyle and Vogeley (2002). The algorithm has been 
extensively employed to analyse void structure and distribution using 
the results from the recently published Data Release 5 (DR5) of the 
Sloan Digital Sky Survey (SDSS) (see, Foster and Nelson 2007). The 
average distance to the third nearest neighbour ($d$) in the sample and 
its standard deviation ($\sigma$) are calculated. In order to ensure 
a high degree of confidence in identifying bona fide voids, we use 
the parameter $R_3 = d + \lambda \ \sigma$ to distinguish wall galaxies 
from field galaxies and set $\lambda = 2$. Wall galaxies are defined 
as those galaxies whose third nearest neighbour is closer than $R_3$. 
All other galaxies are field galaxies. The wall galaxies are placed 
in a grid whose basic cell geometry is cubic having a side of length 
$R_3/2$. The empty cells are then identified and each empty cell acts 
as a seed from which holes are grown. A hole is defined as a sphere 
that is entirely devoid of wall galaxies. Its radius and centre are 
computed such that there are exactly 3 wall galaxies on its surface. 
Voids are then formed by amalgamating the overlapping holes starting 
with the largest holes. Only holes whose radius exceeds a certain 
threshold value ($R_{min} = 7.5\,h^{-1}$\,Mpc for this analysis) can 
form voids; those that are smaller are used to map out the boundary 
surface of a pre--existing void. Thus if there are no holes whose size 
exceeds the threshold, no voids will be identified. The position of 
the centre of each void is calculated by finding the ``centre of 
volume". The position of the centre and the volume are calculated 
using Monte Carlo methods and the equivalent spherical radius is 
determined. In the following, this void finder and its results are 
referred to as {\it Foster/Nelson}.

\subsubsection{Gottl\"ober: Empty spheres in point set}

The void finder starts with a selection of point--like objects in
three--dimensional space. These objects can be halos above some  
mass (or circular velocity) or galaxies above some luminosity. 
Thus voids are characterised by the threshold mass or luminosity. 

For the data used here, $N_g = 380$ grid cells in each dimension 
were used, which corresponds to a grid cell size of 158\,$h^{-1}$\,kpc.
On this grid the point is found, which has the largest distance to 
the set of points defined above. This grid point is the centre of the 
largest void. This void is then excluded, and the procedure is repeated
by searching for a point with the largest distance to the set. 
Iterating this procedure thus yields the full sample of voids 

In principle, the algorithm allows to have a certain number of points 
(objects above the threshold mass or luminosity) inside the void. 
Here, this number is set to zero, i.e. the voids are completely empty 
with respect to the defined sample. Of course, they may contain objects 
with smaller masses or lower luminosities than the assumed threshold.

In principle, the algorithm allows for the construction of voids with
arbitrary shape. The starting point is the spherical void described
above. It can be extended by spheres of lower radius which grow from
the surface of the void into all possible directions. However, in this
test case this feature was switched off, and the search was restricted
to spherical voids to avoid ambiguities of the definition of allowed 
deviations from spherical shape. In the following, this void finder and 
its results are referred to as {\it Gottl\"ober}.

\subsubsection{Hoyle/Vogeley: Voidfinder}

{\it Voidfinder} was introduced in Hoyle \& Vogeley (2002; {\it HV02}) and has 
been used frequently to locate voids in galaxy surveys (Hoyle \& Vogeley 
2004, Hoyle et al. 2005). Full details of how {\it voidfinder} works can be 
found in Hoyle \& Vogeley (2002), so here we will only briefly summarise 
the algorithm. {\it Voidfinder} operates on samples of galaxies and is 
based on the ideas discussed in El--Ad \& Piran (1997) and El--Ad et 
al. (1997). In a volume--limited galaxy catalogue (with a typical limit
just fainter than M*), galaxies are first pre--categorised into wall or 
void galaxies, depending on the distances to the galaxies' third--nearest 
neighbours. Wall galaxies are then binned into cells of a cubic grid. 
Around the centres of all empty grid cells the largest possible spheres 
that are also empty are found. Finally, the set of unique voids is 
constructed by determining maximal spheres and their overlaps. 
{\it voidfinder} voids are non--spherical. A minimum void size of 
$10\,h^{-1}\,$Mpc is set to only select the largest, statistically
most significant voids. For tests of {\it voidfinder} using simulation
data see Benson et al. (2003). In the following, this void finder and its 
results are referred to as {\it Hoyle/Vogeley}.

\subsubsection{M\"uller: Empty convex regions in point set}

This grid based void finder looks first for empty base voids in the 
halo/galaxy sample, and then it adds extensions to approximate spherical 
voids. It was run first with only a base void search and then with extensions.
The idea of the void finder is to look for empty nearly convex regions
in the galaxy distribution. It is based on a grid on the survey volume
where cells with galaxies are marked as occupied. In the next step,
it looks for maximum cubes on the grid that are empty of galaxies and
previously found voids. We call this a base voids, and get a first
catalogue of cube voids, the most simple algorithm, but it produces a
void catalogue with similar sizes as assuming a spherical base volume.
A slightly more refined method is to extend the base voids along the
faces with adding square sheets empty of galaxies and not contained in
previously found voids. This extension procedure is iterated. To avoid
extended fingers or bridges between voids, we require the extension to
have a surface bigger than $2/3$ (an arbitrary parameter) of the previous
one. These extended voids have on average the same volume in the
extensions as in the base voids. We measure the size by the effective
cube size, i.e. a cube of the same volume as the base plus extensions.
Such voids are in general larger than selecting square voids. This
void finder is based on the prescription of Kauffmann \& Fairall
(1991), which was further developed and tested by M\"uller et al. (2000) 
and Arbabi-Bidgoli \& M\"uller (2002). It uses a $300^3$ grid and 
includes voids with a minimum effective radius of $3\,h^{-1}$\,Mpc.
In the following, this void finder and its results are referred to 
as {\it M\"uller}.

\begin{table*}
\begin{tabular}{|lllllllllll|}
\hline
{\bf Author} & & $N_V$ & $FF_V$ & $\delta_{\mbox{\scriptsize{DM}}}$ & $N_g$ & 
       $\delta_{\mbox{\scriptsize{g}}}$ & $N_{g,20}$ & 
       $\delta_{\mbox{\scriptsize{g,20}}}$ & ($x_{\mbox{\scriptsize{max}}}$,  $y_{\mbox{\scriptsize{max}}}$, 
               $z_{\mbox{\scriptsize{max}}}$)  & $r$ \\ 
             & & & & & & & & & [Mpc/$h$] & [Mpc/$h$] \\ 

\hline
Brunino               & P  &   3 & 0.37 & -0.78 &  754 & -0.71 &  7 & -0.93 & (38.6, 46.8, 199.5) & 16.0 \\
Colberg$^1$           & DM &  21 & 0.92 & -0.74 & 2258 & -0.65 & 35 & -0.85 & (35.3, 41.2, 193.9) & 29.9 \\
Fairall               & P  &  18 & 0.59 & -0.73 & 1376 & -0.67 & 25 & -0.83 & (33.0, 40.0. 200.0) & 20.0 \\
Foster/Nelson         & P  &   3 & 0.41 & -0.82 &  114 & -0.96 &  0 & -1.00 & (36.3, 36.6, 192.4) & 18.0 \\
Gottl\"ober$^2$       & P  &   9 & 0.35 & -0.77 &  733 & -0.70 &  0 & -1.00 & (32.1, 44.0, 192.0) & 16.4 \\
Hahn/Porciani$^{1,3}$ & DM &  14 & 0.29 & -0.73 &  248 & -0.92 &  0 & -1.00 & (30.5, 33.6, 191.8) & 17.2 \\
Hoyle/Vogeley$^{1,2}$ & P  &   4 & 0.84 & -0.68 & 2166 & -0.56 & 40 & -0.79 & (31.9, 47.1, 193.2) & 24.6 \\
M\"uller$^2$          & P  &  24 & 0.58 & -0.76 & 1469 & -0.65 &  0 & -1.00 & (30.7, 42.7, 189.1) & 25.6 \\ 
Neyrinck$^{1,3,4}$    & DM &  29 & 0.32 & -0.68 &  834 & -0.63 & 14 & -0.83 & (30.3, 33.5, 194.9) & 11.3 \\
Pearce                & DM &   5 & 0.15 & -0.90 &   51 & -0.95 &  0 & -1.00 & (35.9, 33.8, 193.5) & 11.9 \\
Platen/Weygaert$^1$   & DM & 167 & 1.0  & -0.91 &   18 & -1.00 &  0 & -1.00 & (37.5, 36.2, 194.3) & 14.3 \\
Plionis/Basilakos     & DM &  15 & 0.13 & -0.92 &    0 & -1.00 &  0 & -1.00 & (37.1, 33.8, 192.7) & 10.0 \\
Shandarin/Feldman     & DM &  19 & 0.23 & -0.88 &    0 & -1.00 &  0 & -1.00 & (31.5, 41.1, 192.7) & 17.1 \\
\hline
\end{tabular}
\caption{An overview of some of the main results of this study: for each void
         finder, we give the total number of voids, $N_V$, in the volume considered
         here, the volume filling fraction, $FF_V$, the average dark matter overdensity,
         $\delta_{\mbox{\scriptsize{DM}}}$, of the voids, the total
         number of galaxies, $N_g$, found in voids, the average galaxy overdensity 
         $\delta_{\mbox{\scriptsize{g}}}$, the number of galaxies  
         brighter than $m_B = -20$, $N_{g,20}$, found in voids, the 
         average galaxy overdensity using only galaxies brighter than $m_B = -20$,
         $\delta_{\mbox{\scriptsize{g20}}}$, and positions of the centres of the largest void and
         their radii. We also classify the void finders into those
         using the dark matter (smoothed or not -- DM) and those using points
         (galaxies or haloes -- P).
         Notes: $^1$ the voids are non--spherical, so the quoted radius is an approximation,
         assuming a spherical void. $^2$ using the $B < -20$ galaxy
         sample. $^3$ the quoted centre of the void is actually the position
         of lowest density. $^4$ 9308 voids found; of them 2362, 525, 164, 64, 29, 13, and 5 
         exceed 1 through 7$\sigma$ probability thresholds,
         respectively. We use the 5$\sigma$ results for comparisons.}
\label{tab:results}
\end{table*}
\begin{figure*}
  \begin{center}
    \begin{tabular}{cc}
      \begin{minipage}{51mm}
        \begin{center}
          \includegraphics[width=51mm]{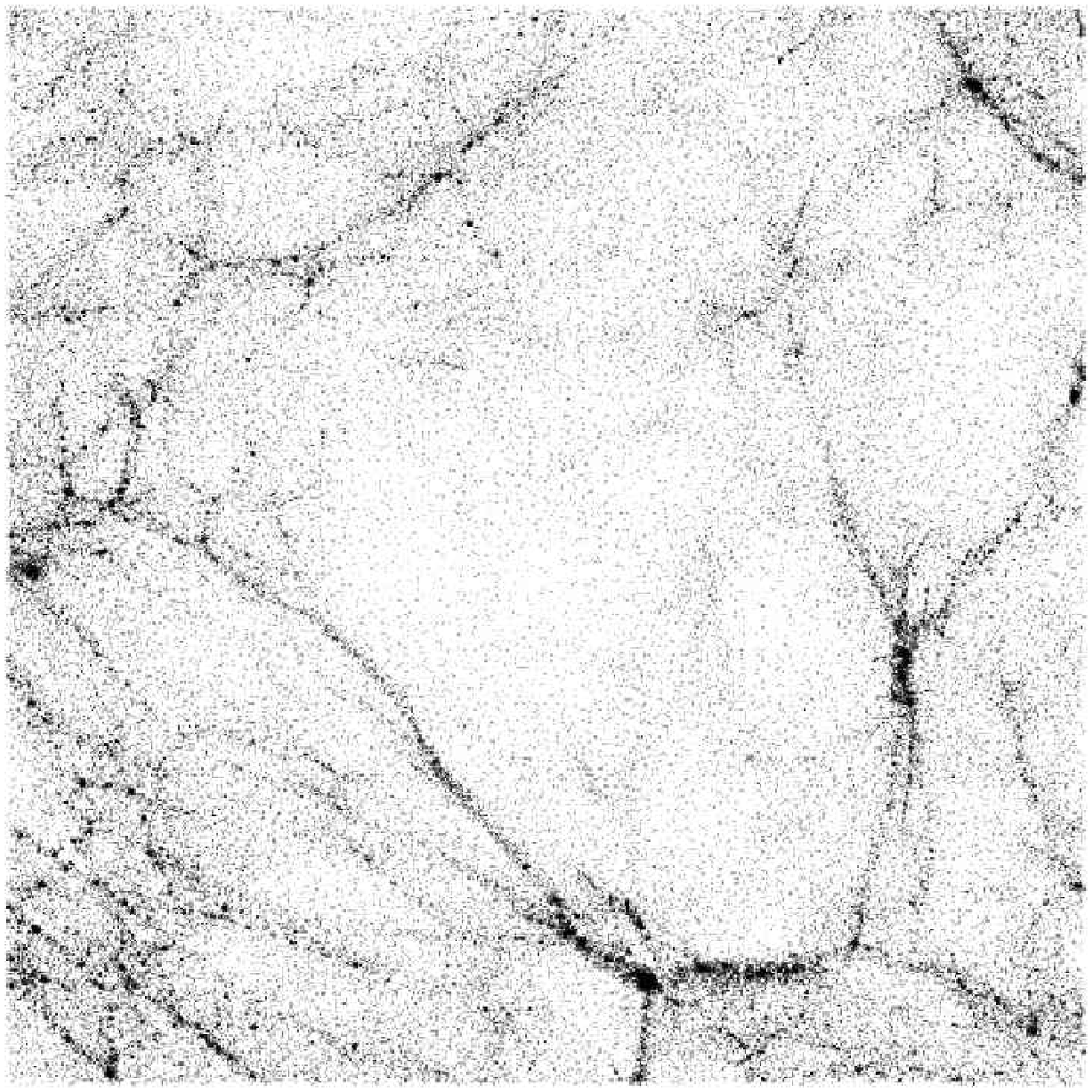}
        \end{center}
      \end{minipage}
      \hspace{0.3cm}
      \begin{minipage}{51mm}
        \begin{center}
          \includegraphics[width=51mm]{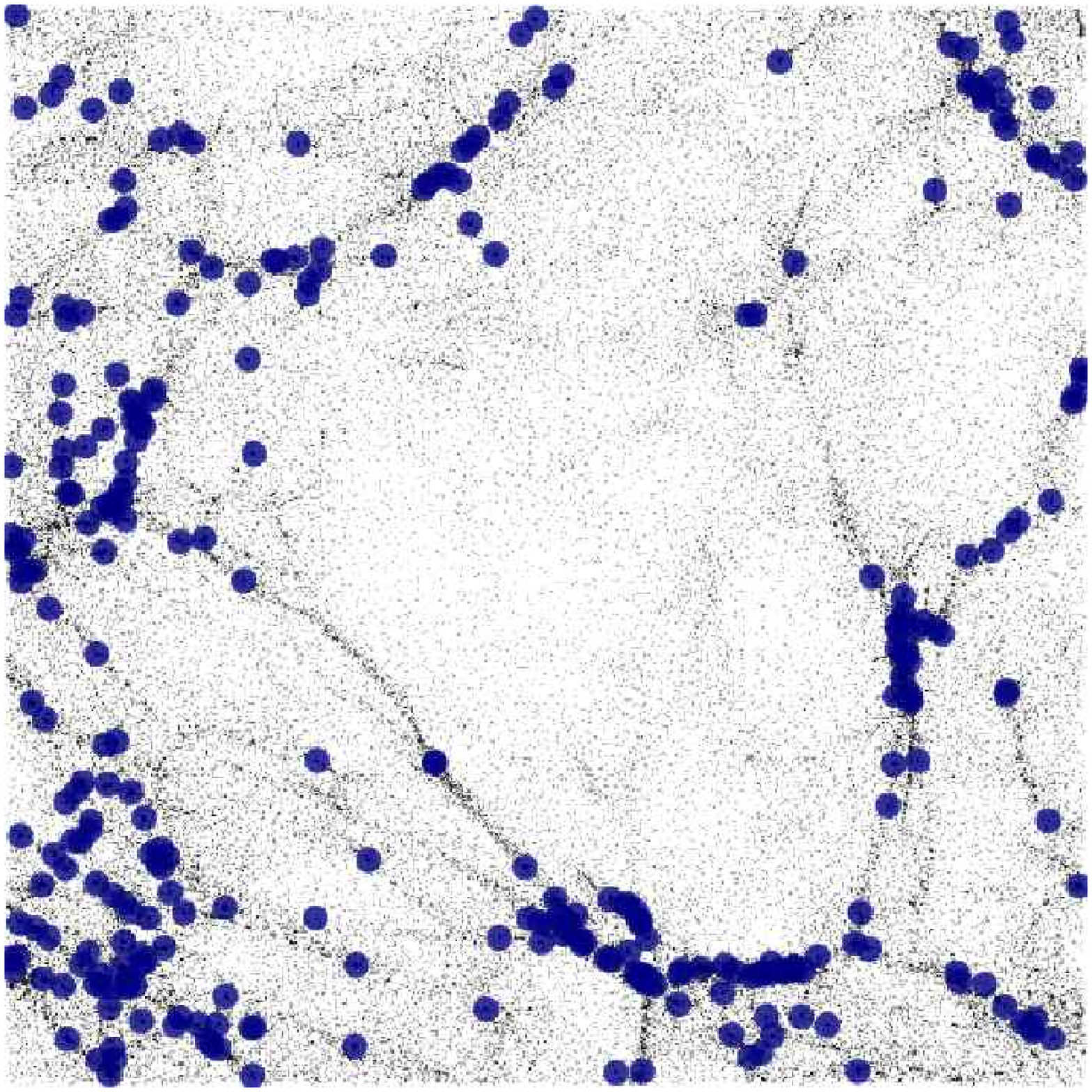}
        \end{center}
      \end{minipage}
      \hspace{0.3cm}
      \begin{minipage}{51mm}
        \begin{center}
          \includegraphics[width=51mm]{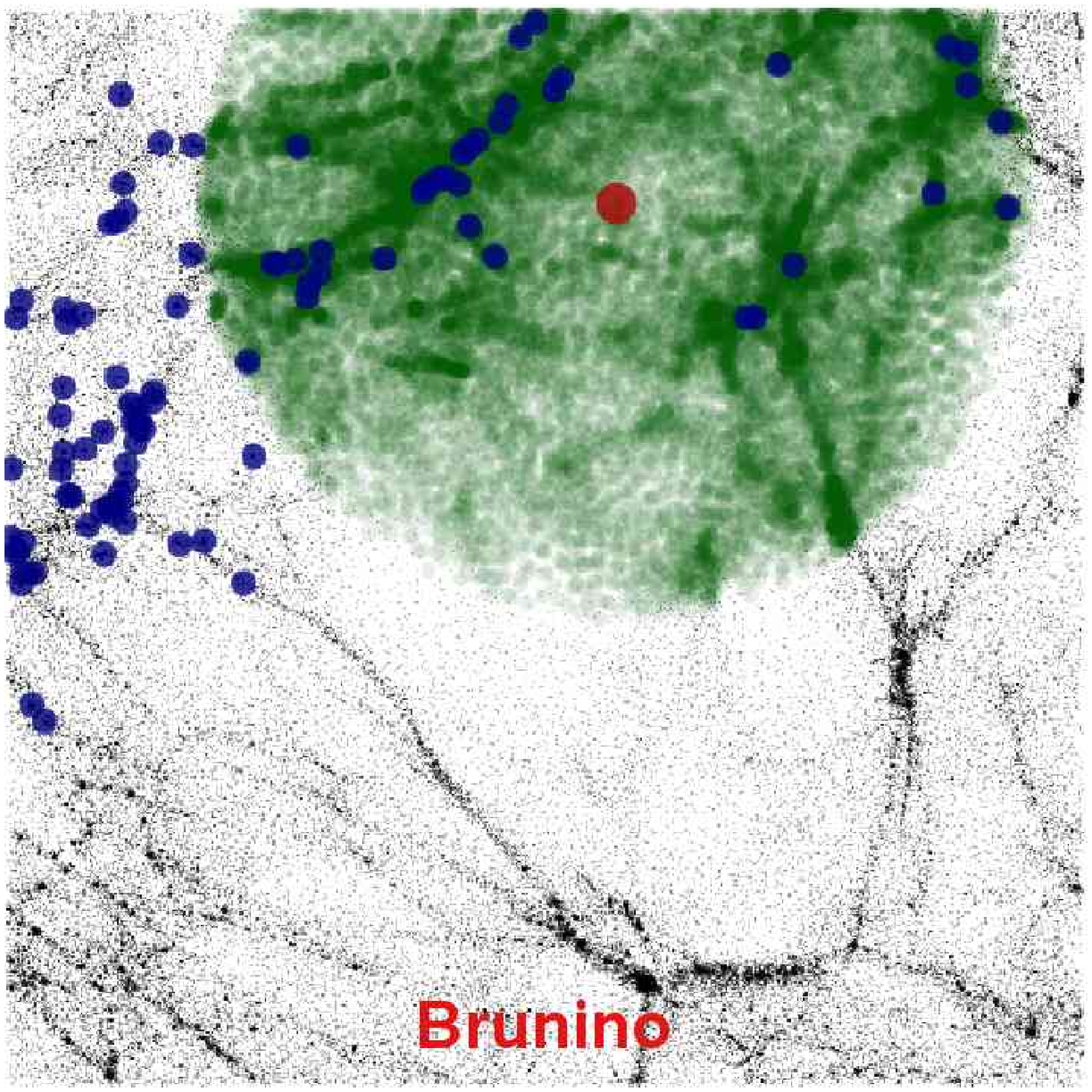}
        \end{center}
      \end{minipage}
    \end{tabular}
    \vspace{2mm}
    \begin{tabular}{cc}
      \begin{minipage}{51mm}
        \begin{center}
          \includegraphics[width=51mm]{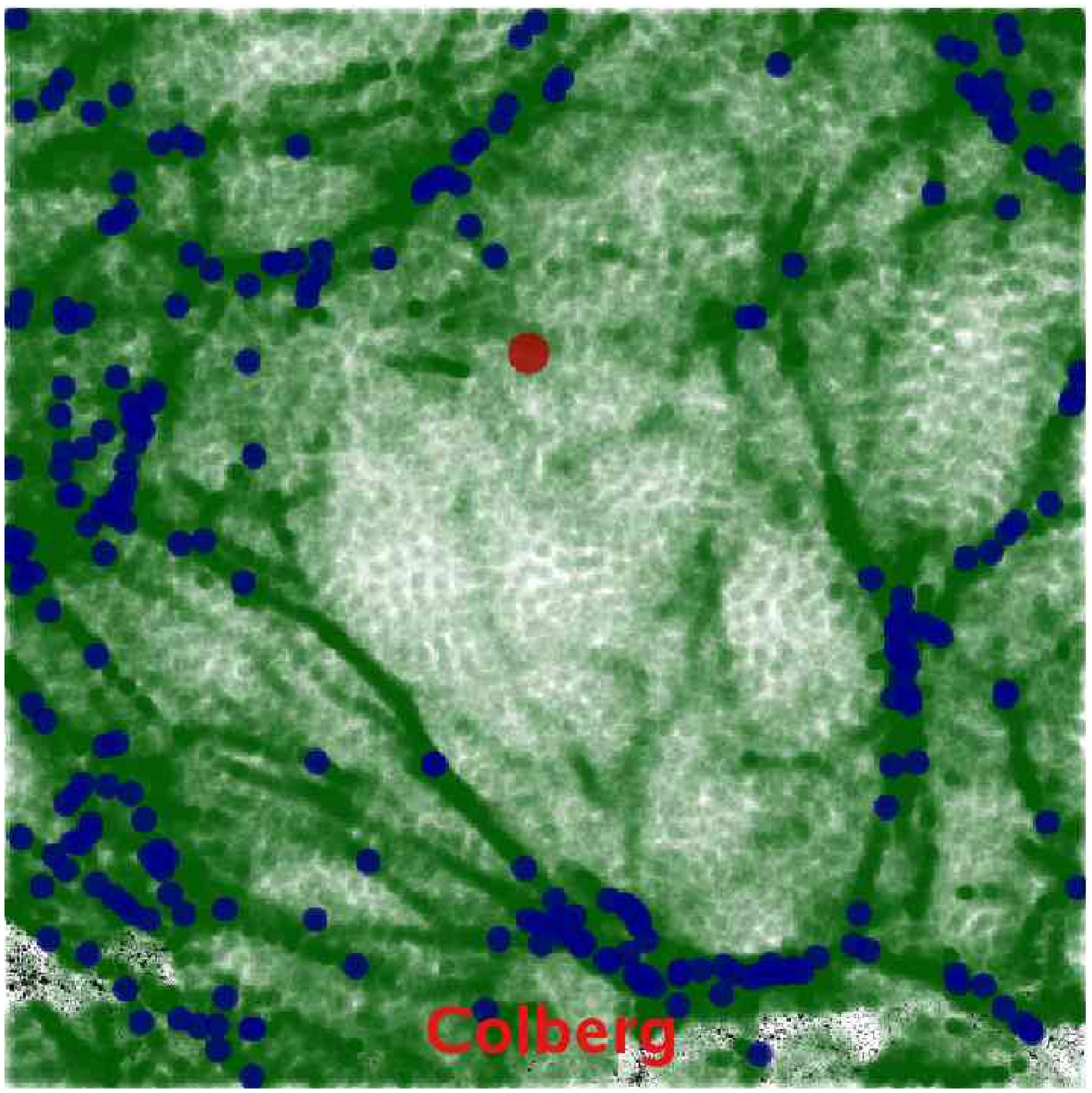}
        \end{center}
      \end{minipage}
      \hspace{0.3cm}
      \begin{minipage}{51mm}
        \begin{center}
          \includegraphics[width=51mm]{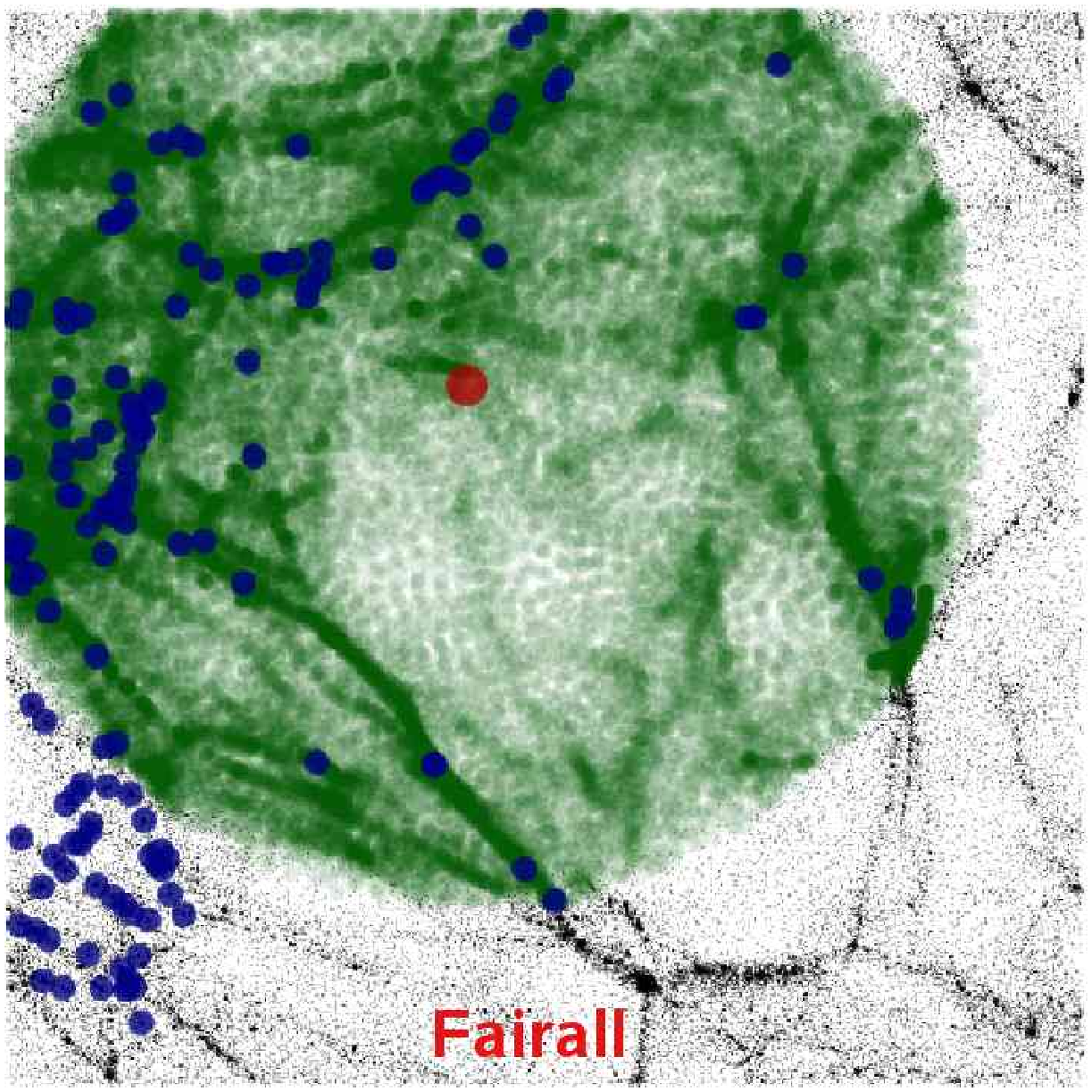}
        \end{center}
      \end{minipage}
      \hspace{0.3cm}
      \begin{minipage}{51mm}
        \begin{center}
          \includegraphics[width=51mm]{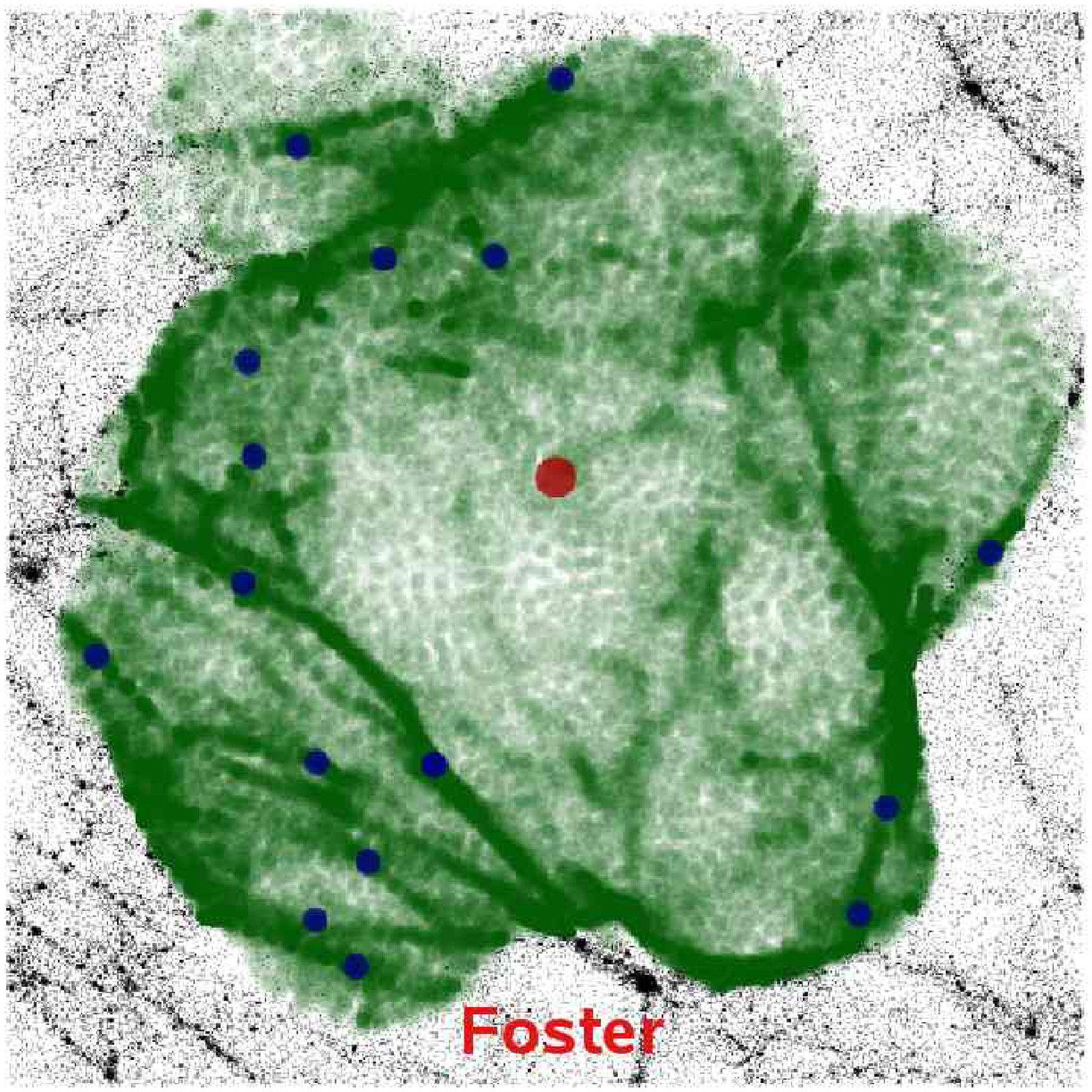}
        \end{center}
      \end{minipage}
    \end{tabular}
    \vspace{2mm}
    \begin{tabular}{cc}
      \begin{minipage}{51mm}
        \begin{center}
          \includegraphics[width=51mm]{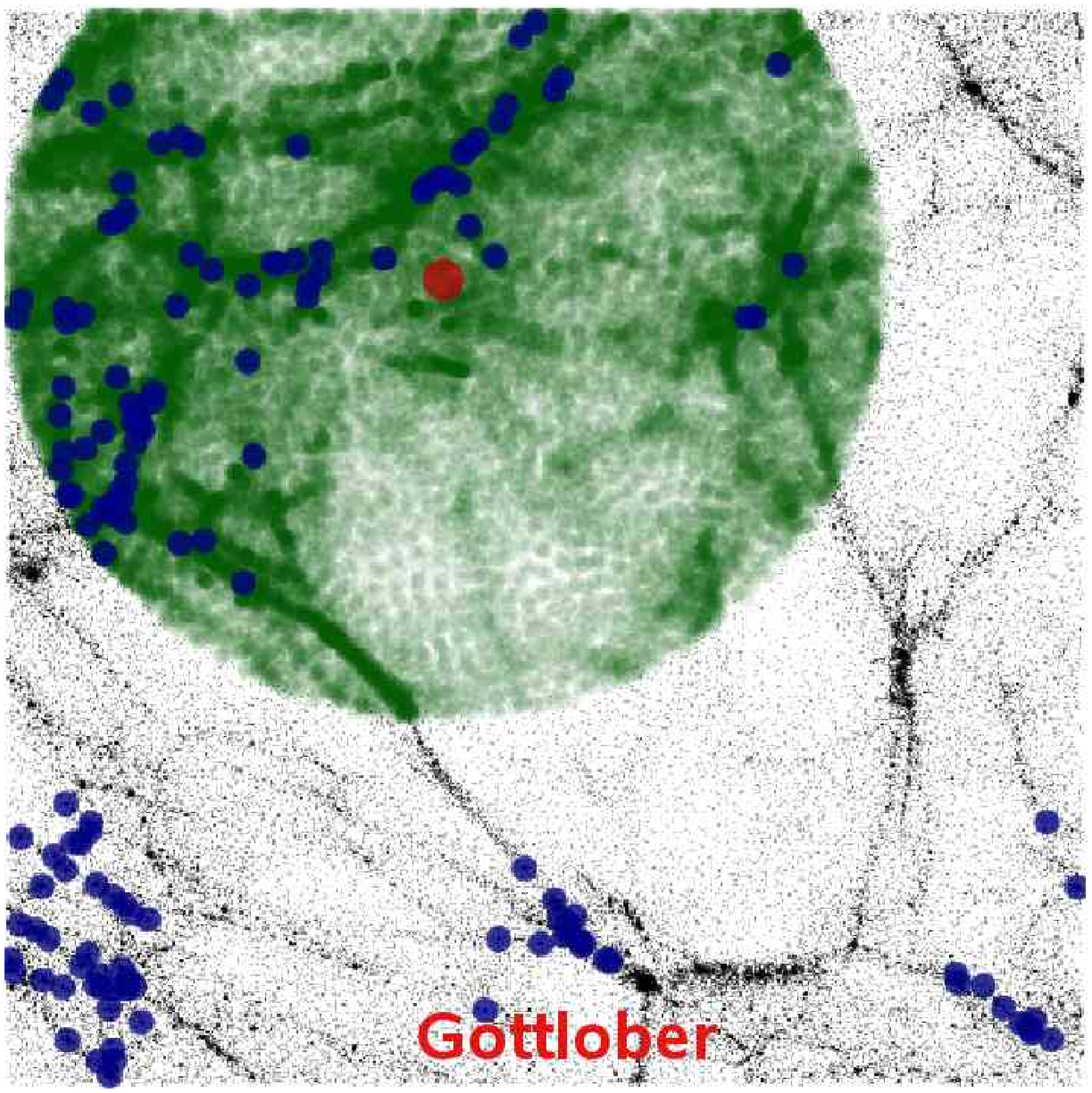}
        \end{center}
      \end{minipage}
      \hspace{0.3cm}
      \begin{minipage}{51mm}
        \begin{center}
          \includegraphics[width=51mm]{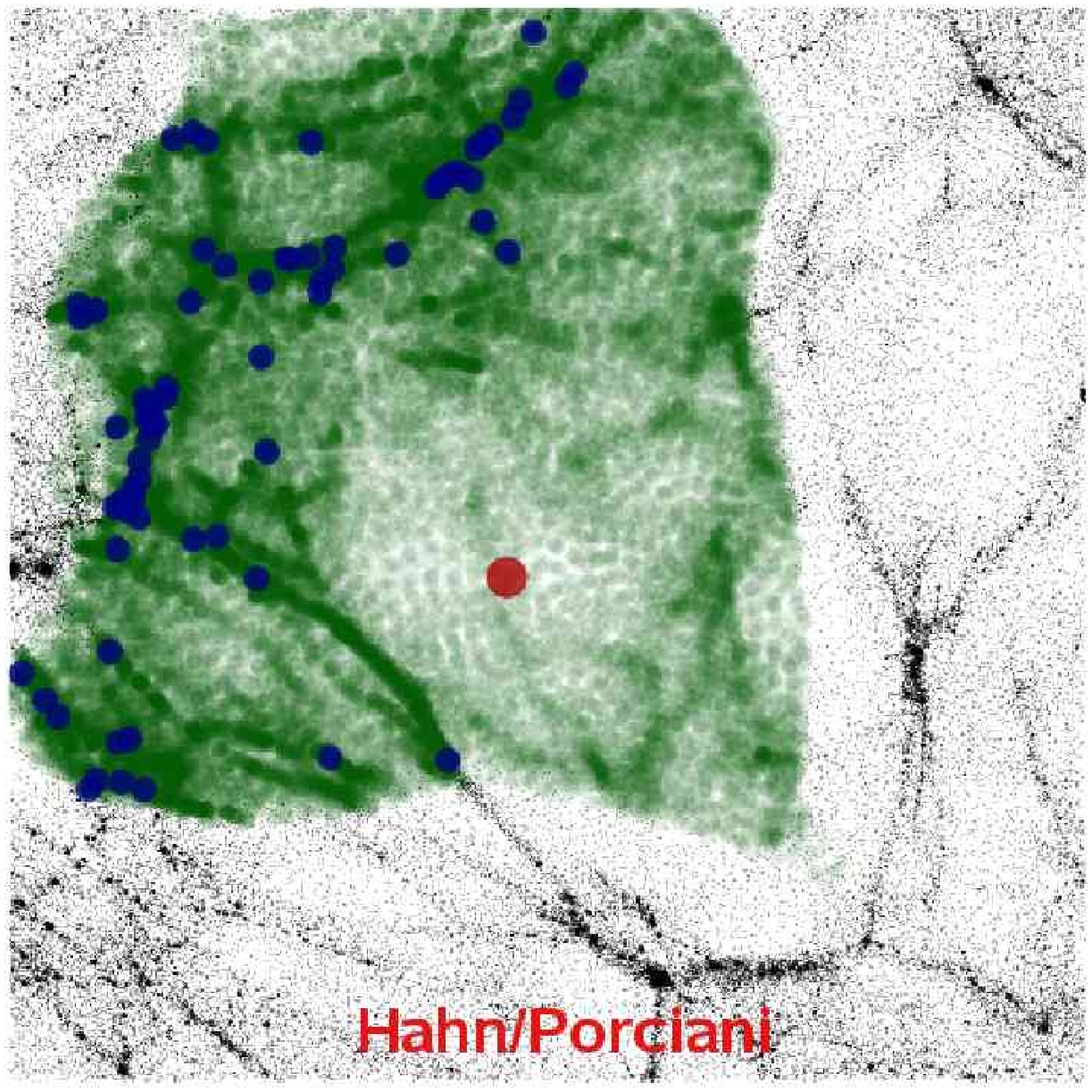}
        \end{center}
      \end{minipage}
      \hspace{0.3cm}
      \begin{minipage}{51mm}
        \begin{center}
          \includegraphics[width=51mm]{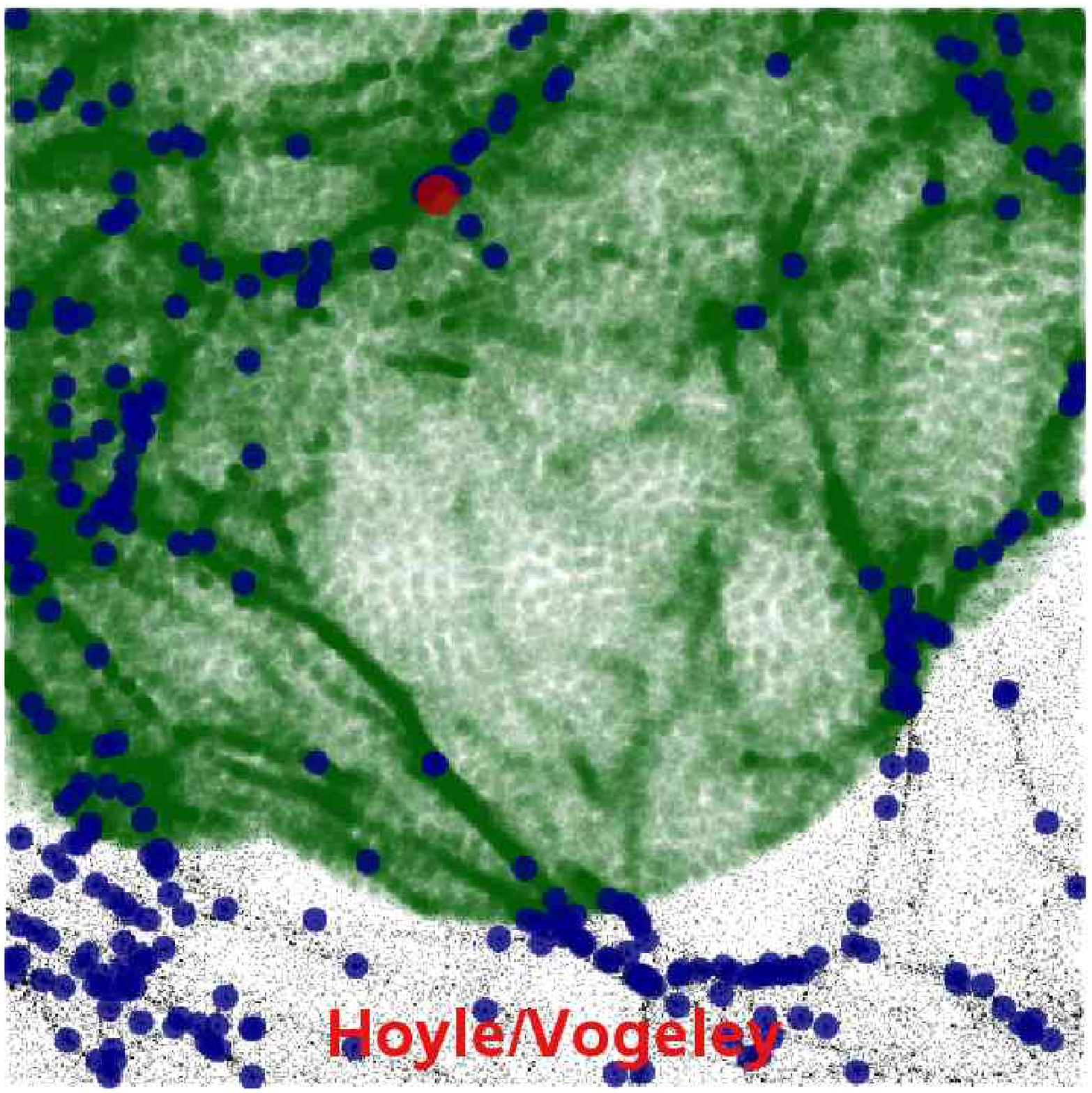}
        \end{center}
      \end{minipage}
    \end{tabular}
    \vspace{2mm}
    \begin{tabular}{cc}
      \begin{minipage}{51mm}
        \begin{center}
          \includegraphics[width=51mm]{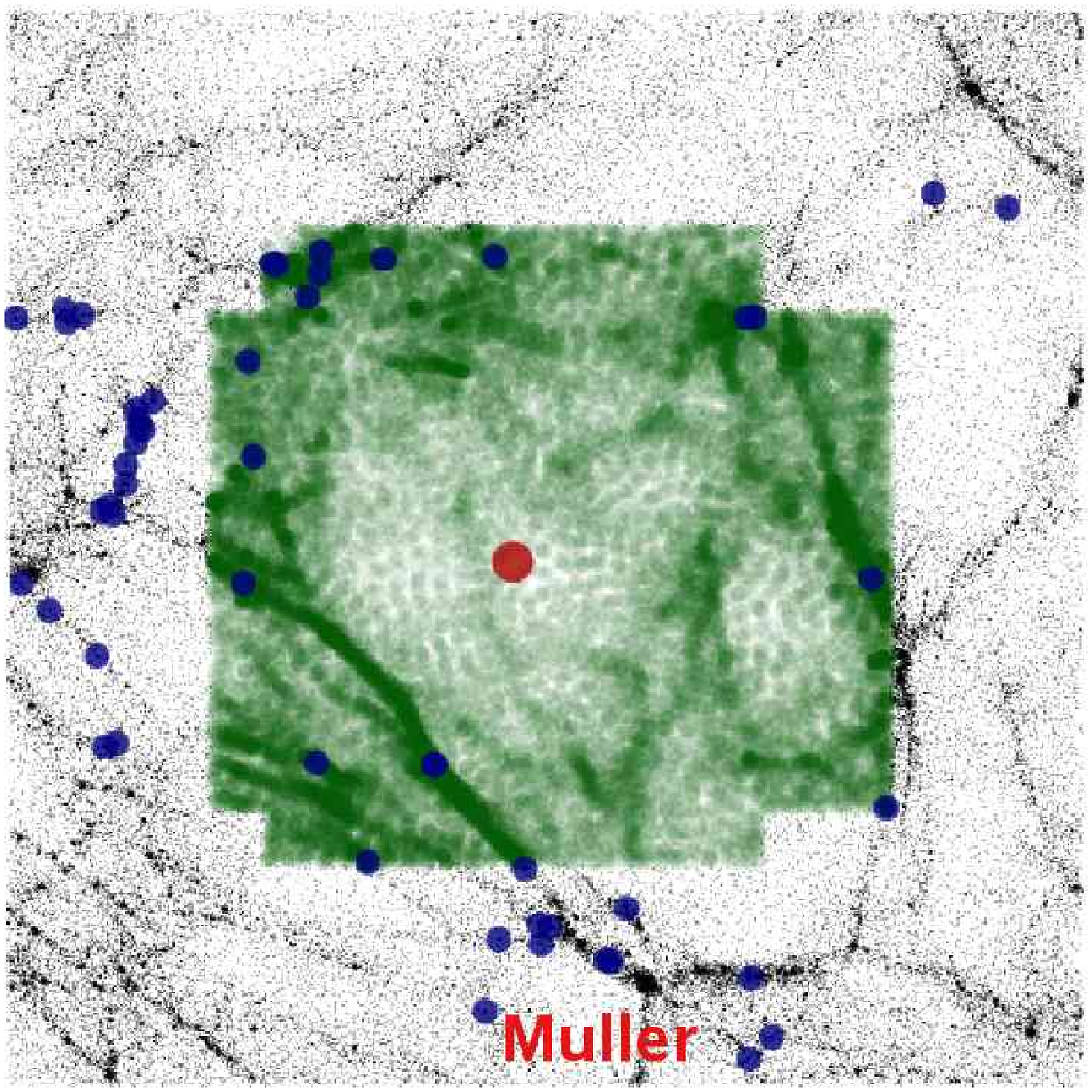}
        \end{center}
      \end{minipage}
      \hspace{0.3cm}
      \begin{minipage}{51mm}
        \begin{center}
          \includegraphics[width=51mm]{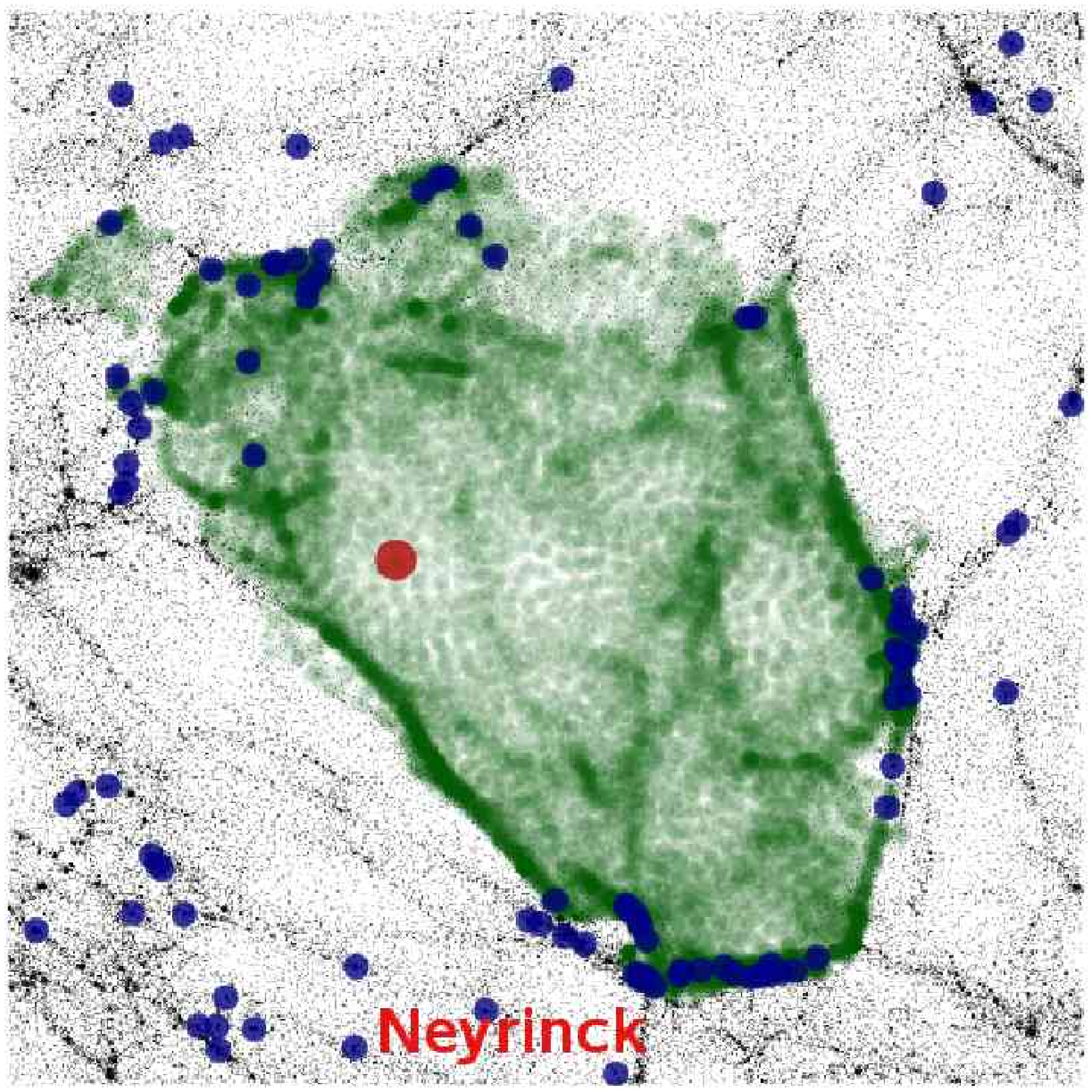}
        \end{center}
      \end{minipage}
      \hspace{0.3cm}
      \begin{minipage}{51mm}
        \begin{center}
          \includegraphics[width=51mm]{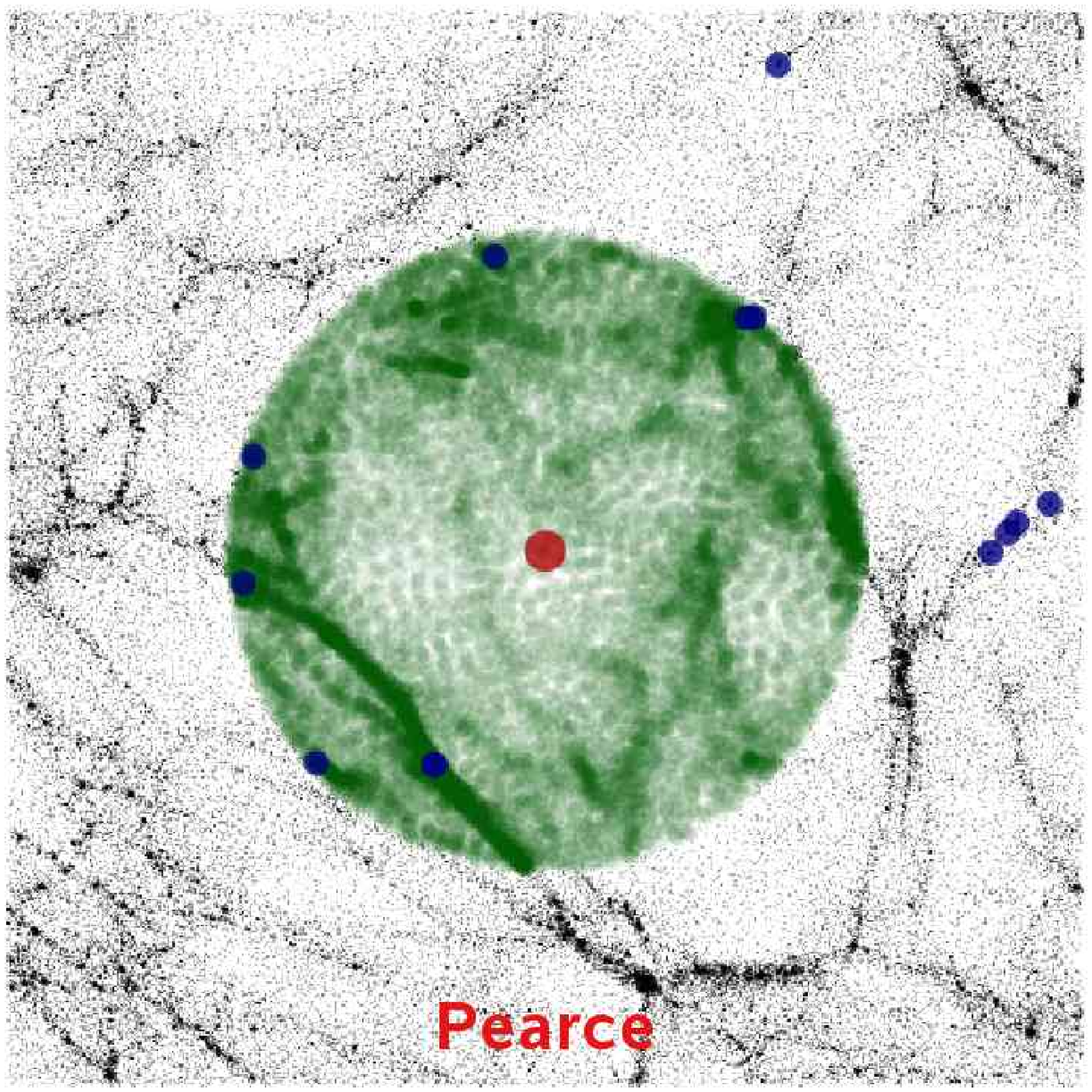}
        \end{center}
      \end{minipage}
    \end{tabular}
  \end{center}
  \caption{A slice of thickness 5\,$h^{-1}$\,Mpc through the 
           centre of the region extracted from the Millennium simulation. 
           The image shows the dark matter distribution 
           in the central 40\,$h^{-1}$\,Mpc region. Void galaxies
           (within any void, not just the largest one) 
           are 
           superimposed on the dark matter distribution as blue circles. The top left and
           top centre panels show only the dark matter distribution and 
           dark matter plus all galaxies in the slice, respectively. 
           The other panels show the locations of the largest void (with dark matter
	   particles inside the void marked green), its centre
	   (red circle), and all void galaxies found by Brunino (top right), 
           Colberg (second row, left column), Fairall (second row, centre), 
           Foster (second row, right column), Gottl\"ober (third row, left
           column), Hahn/Porciani (third row, centre), Hoyle/Vogeley (third 
           row, right column), M\"uller (bottom, left column), Neyrinck 
           (bottom, centre), Pearce (bottom, right column).}
  \label{fig:void_gals1}
\end{figure*}

\begin{figure*}
  \begin{center}
    \begin{tabular}{cc}
      \begin{minipage}{51mm}
        \begin{center}
          \includegraphics[width=51mm]{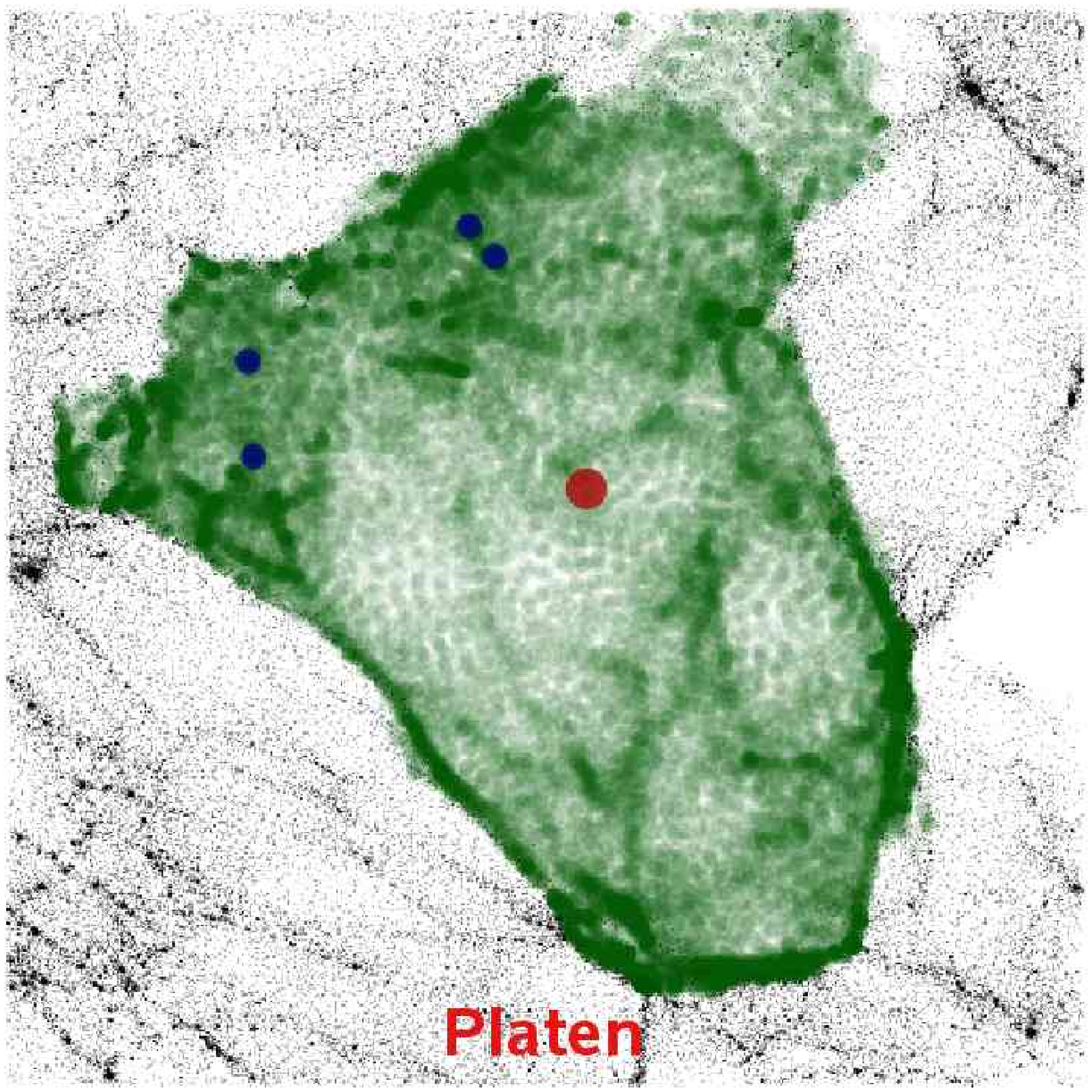}
        \end{center}
      \end{minipage}
      \hspace{0.3cm}
      \begin{minipage}{51mm}
        \begin{center}
          \includegraphics[width=51mm]{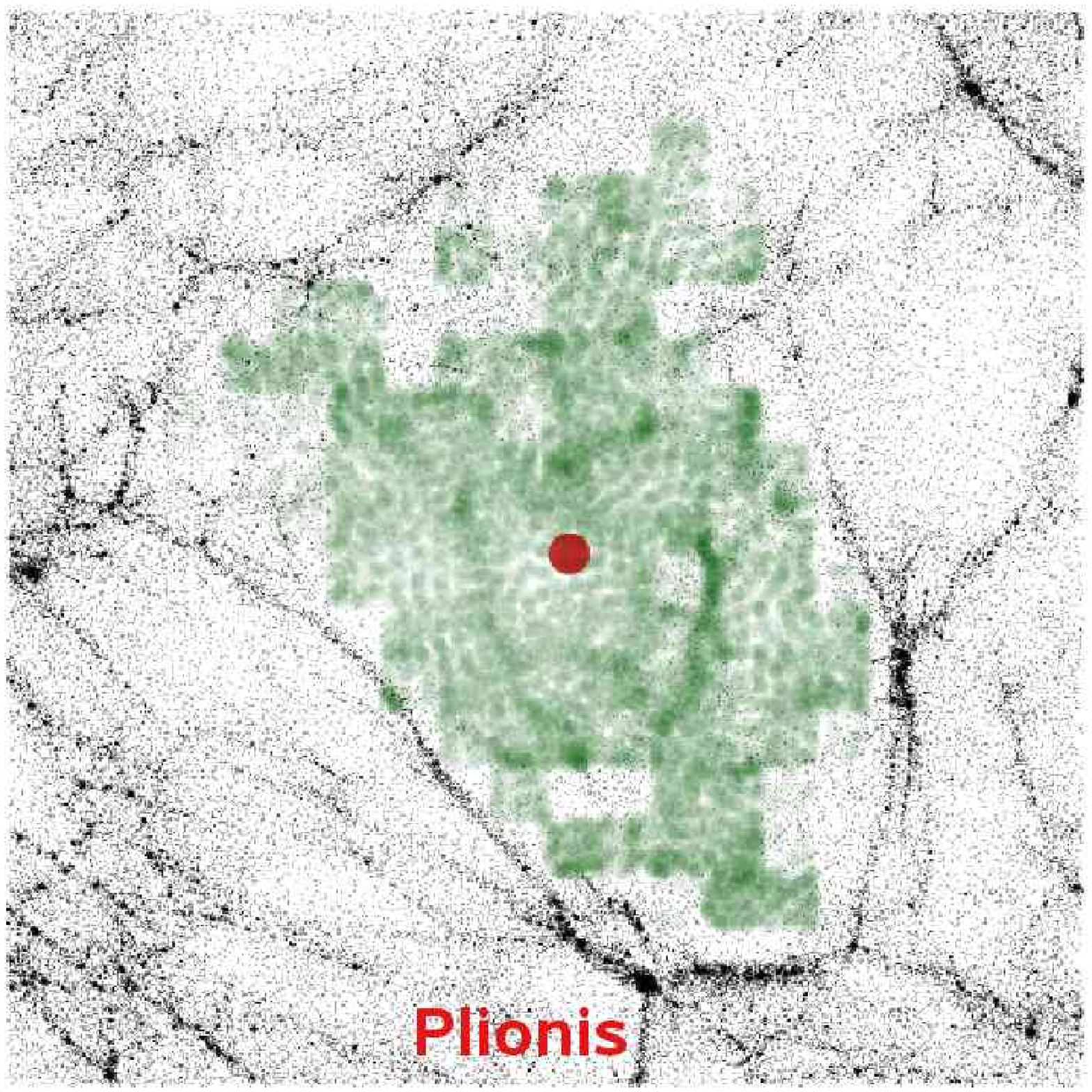}
        \end{center}
      \end{minipage}
      \hspace{0.3cm}
      \begin{minipage}{51mm}
        \begin{center}
          \includegraphics[width=51mm]{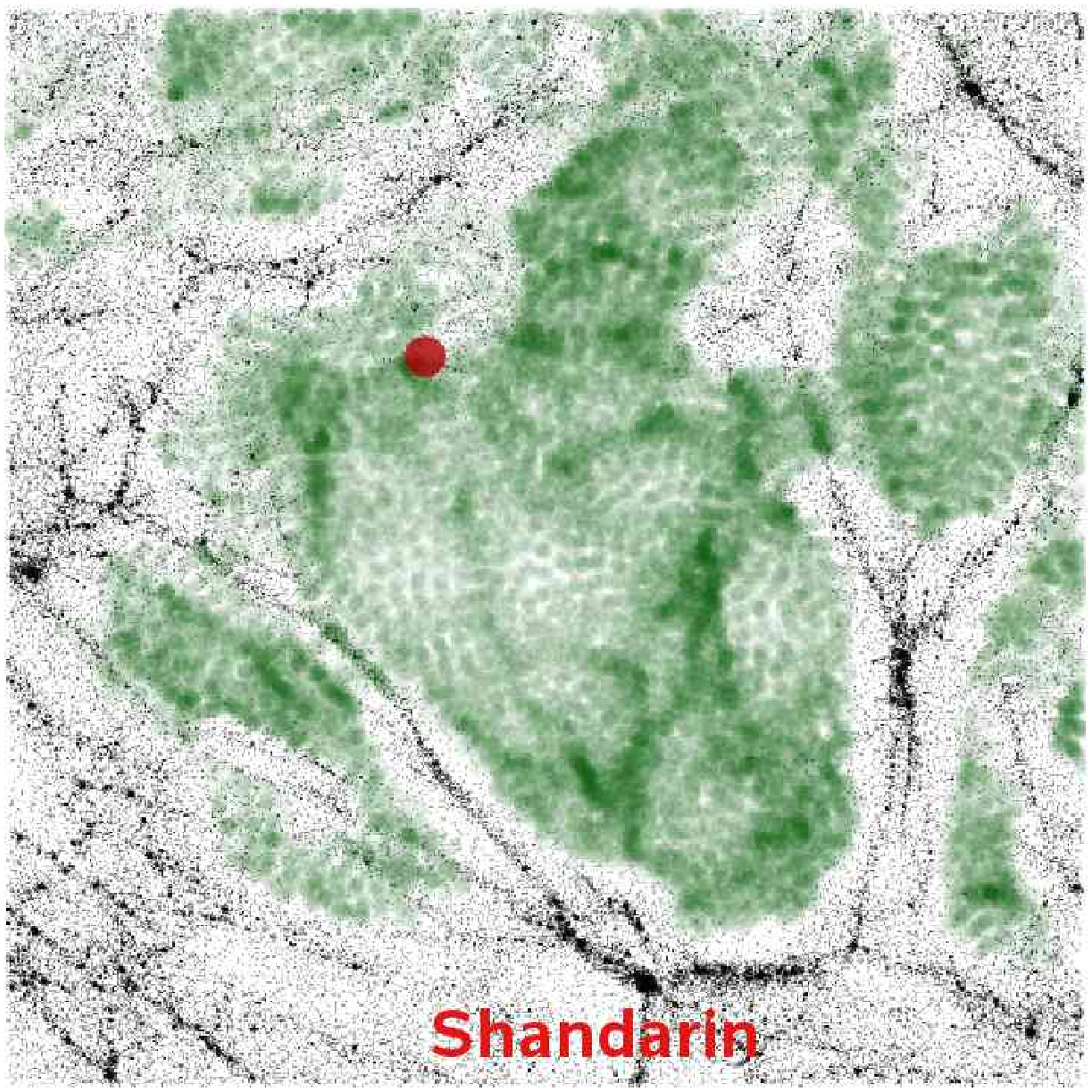}
        \end{center}
      \end{minipage}
    \end{tabular}
  \end{center}
  \caption{Same as and continued from Figure~\ref{fig:void_gals1}.
           Platen/Weygaert (left column), Plionis/Basilakos (centre), and
           Shandarin/Feldman (right column). 
           Note that both Plionis/Basilakos
           and Shandarin/Feldman find no void galaxies.}
  \label{fig:void_gals2}
\end{figure*}

\begin{figure*}
\includegraphics[width=170mm]{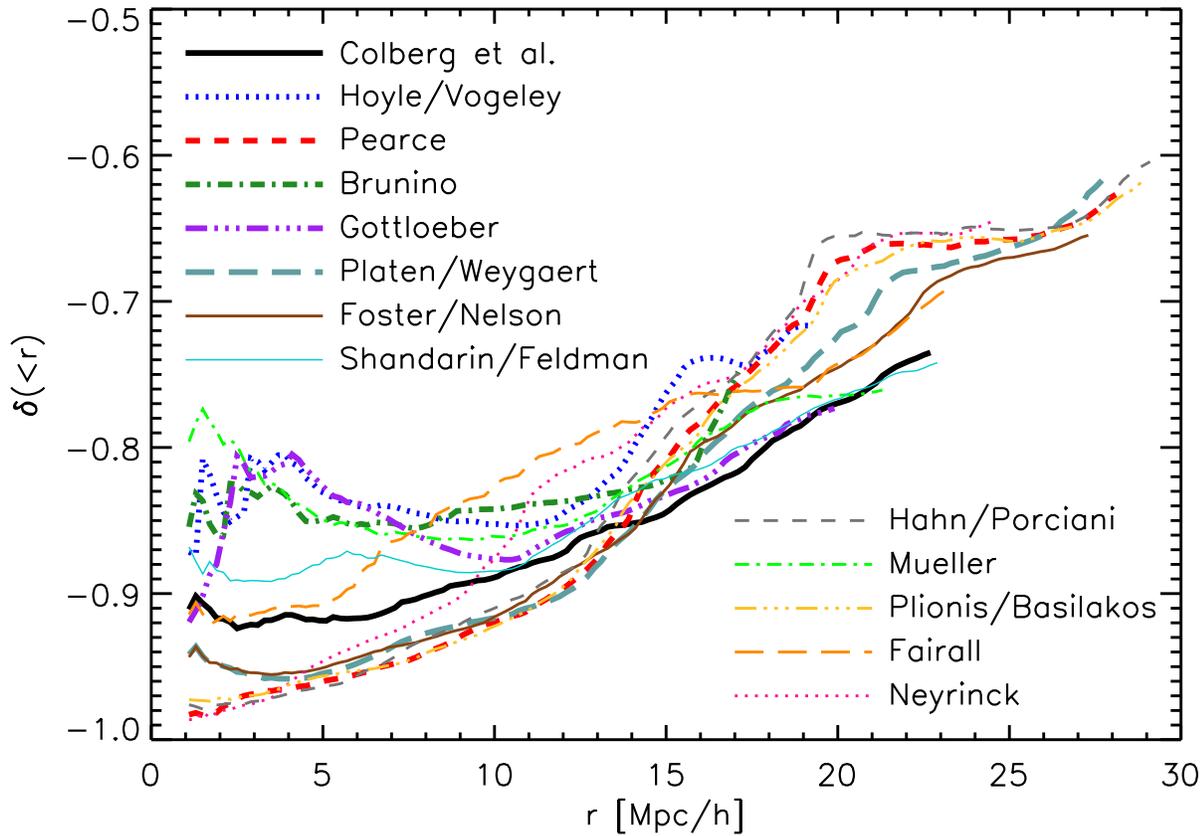}
\vspace{-0.2cm}
\caption{Radially averaged dark matter density profiles of the largest 
         void in each of the void catalogues found by the groups involved 
         in the study. For each void finder the profile extends out to the
	 largest radius that can be studied, given the size of the volume.
	 See main text for more details.}
\label{fig:dens_prof}
\end{figure*}

\begin{figure*}
\includegraphics[width=170mm]{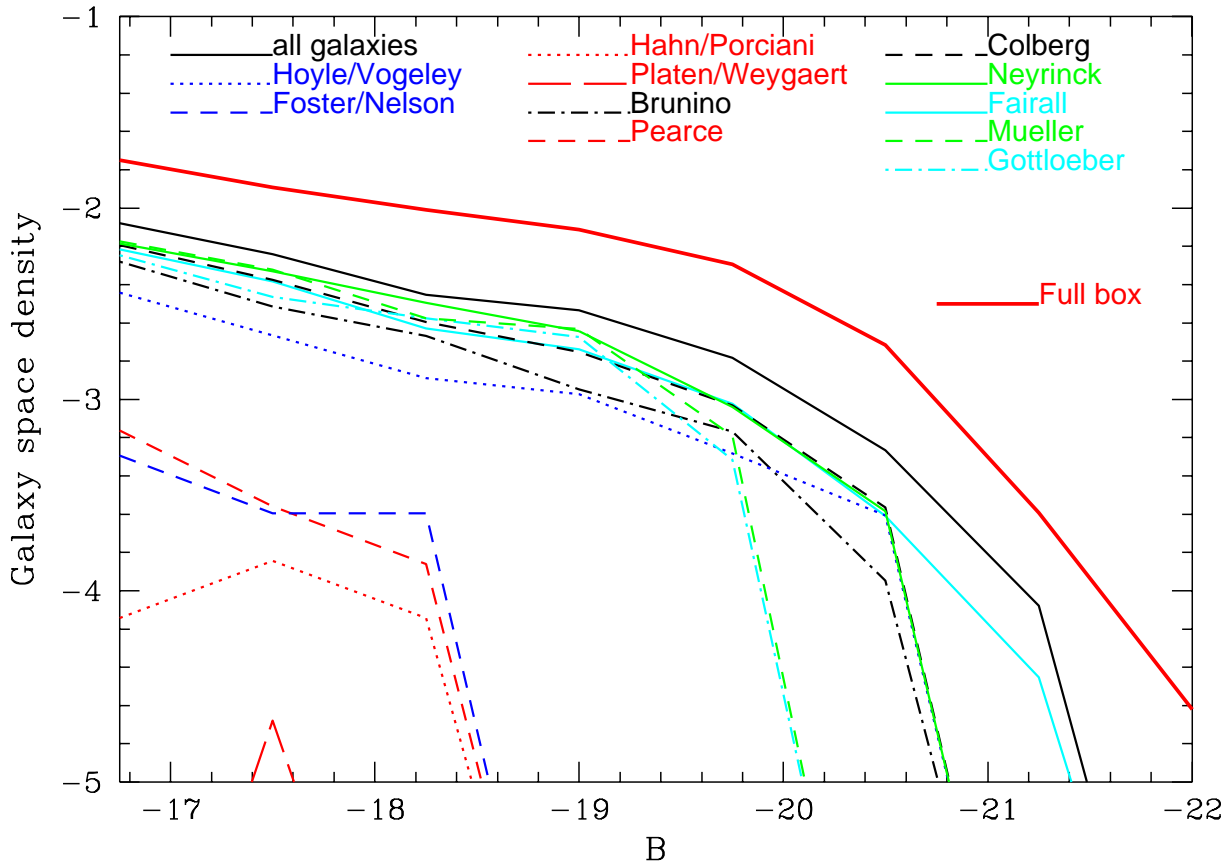}
\vspace{-5.0cm}
\caption{Space density of galaxies (h$^3$/Mpc$^3$/mag) as a function
         of dust corrected $M_B$ for galaxies in the volume under consideration
         and in the catalogues of those void finders which identify galaxies
         inside voids. For purposes of comparison, the luminosity function of
         the full simulation volume is also given. Each void finder
         luminosity function is corrected for the volume occupied by
         the relevant void sample.}
\label{fig:LF}
\end{figure*}

\begin{figure*}
\includegraphics[width=170mm]{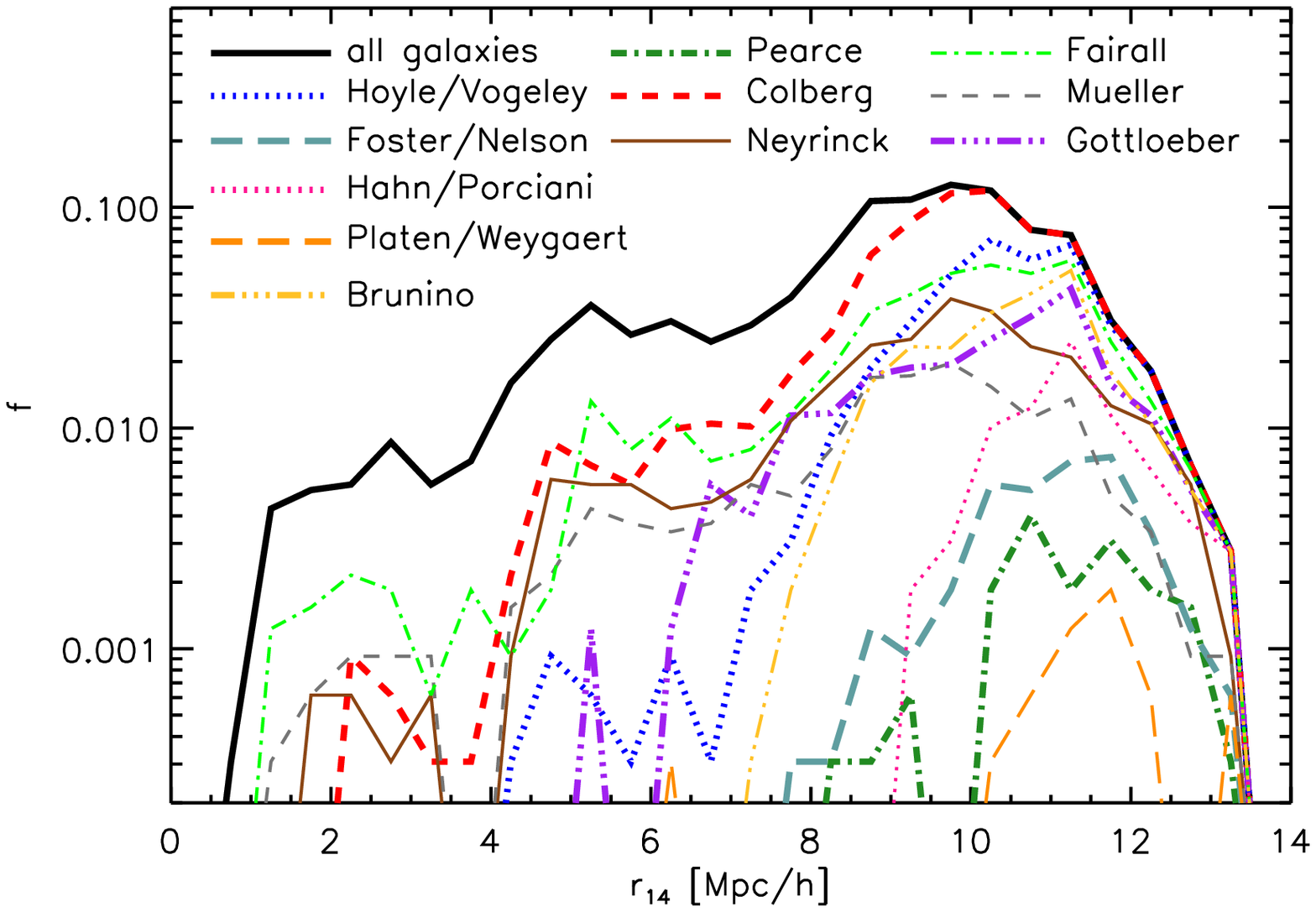}
\vspace{-0.2cm}
\caption{Distributions of the local densities of the galaxies in the results
         of those void finders that identity void galaxies. The local density
         is expressed via $r_{14}$, which for each galaxy gives the radius of 
         the sphere around the galaxy that contains 
         $10^{14}\,h^{-1}$\,M$_\odot$. For comparison purposes, the
         distribution of the full galaxy sample is also shown.}
\label{fig:local_dens}
\end{figure*}

\section{Results -- Comparison} \label{comparison}

\subsection{Basic numbers} \label{sec:numbers}

In Table~\ref{tab:results} we provide an overview of the results obtained
with the different void finders. In particular, for each void finder, we list 
the total number of voids, $N_V$, the volume filling fraction\footnote{The
volume filling fraction is the fraction of the volume that is contained in 
voids, $FF_V = \sum V_i/V_{\rm total}$, where the sum is over all voids in
the sample, and $V_{\rm total}$ is the total volume; so, for example, 
$FF_V = 0.5$ means that voids fill half the volume.}, 
$FF_V$, the average dark matter overdensity, $\delta_{\mbox{\scriptsize{DM}}}$, of the 
voids, the total number of galaxies, $N_g$, found in voids, the corresponding 
average galaxy overdensity, $\delta_{\mbox{\scriptsize{g}}}$, the number of 
galaxies brighter than $m_B = -20$, $N_{g,20}$, found in voids, the 
corresponding average galaxy overdensity using only those galaxies, 
$\delta_{\mbox{\scriptsize{g20}}}$, and positions of the centres of the 
largest void and their radii.

When comparing these numbers it is important to keep the differences in
the void finders in mind. For example, some void finders construct strictly
spherical voids, whereas others build larger ones out of spherical
proto--voids. In addition, there are differences in the spatial resolutions.
The numbers of voids found in the volume thus can be expected to be 
different, and they should merely be treated as illustrative quantities.

If the different results strictly reflected the density field in the
simulation, that is if all the void samples were centred on the most
underdense regions and then extended out to higher density regions, there
would be a simple relationship between the volume filling fraction $FF_V$
and the average dark matter overdensity $\delta_{\mbox{\scriptsize{DM}}}$.
To a certain degree such a correlation does exist. For example, the 
{\it Pearce} voids are centred on the particles with the lowest local
densities and are cut off at an overdensity of $\delta = -0.9$, whereas
{\it Colberg} voids are constructed around proto--voids with $\delta = -0.8$.
This results in a much lower value of $FF_V$ for {\it Pearce}, whereas
{\it Colberg}'s voids fill almost the entire volume\footnote{Recall that the
subvolume studied here has a mean overdensity of $\delta = -0.28$. So
{\it Colberg}'s result is not all that surprising given the procedure it
uses and the fact that the whole region is quite underdense.}. 

A more detailed examination of Table~2 reveals the key difference
between the void finders we have employed: they effectively target
different mean overdensities. Those which correspond to a low mean
overdensity ({\it Pearce, Platen/Weygaert, Plionis/Basilakos \& Shandarin},
all with $\delta \sim -0.9$) naturally contain very few galaxies as
they pick out the deepest parts of the voids. At the other extreme,
those finders which effectively employ higher overdensity thresholds
(for instance {\it Colberg} and {\it Hoyle/Vogeley} with $\delta \sim -0.7$) pick
out much larger regions and naturally enclose far more galaxies. There
is nothing intrinsically wrong with different void finders targeting
different overdensities, in fact in some sense pretty much the whole
region could be classed as a ``void'', in that it has far
less dark matter than expected and consequently is depleted of galaxies.
As a result secondary characteristics such as the void
radius or the number density of void galaxies need to be calibrated
against this effective threshold before techniques can be compared in
detail. Such a study is beyond the scope of this paper
but should be borne in mind when examining such measures as the
largest void in any particular dataset.


Given that not all void finders are density--based, there also is no direct
relationship between $FF_V$ and the number of galaxies inside the voids,
$N_g$. There is a clear difference in $N_g$ between the different models,
with {\it Plionis/Basilakos} and {\it Shandarin/Feldman} finding no 
void galaxies whatsoever\footnote{Note, though that for {\it Plionis/Basilakos} a
change of the pdf fraction, below which ``void cells'' are considered, to
the lowest 30\% of the pdf, results in finding void galaxies.}, 
and the most extreme cases with several thousand void galaxies. 
Given the fact that the volume studied here has a mean overdensity of
$\delta = -0.28$ finding lots of void galaxies is maybe not all that 
surprising -- provided one is happy with the existence of such objects.
The number of galaxies brighter than $m_B = -20$, $N_{g,20}$, is either
zero or very small for all void finders. This is an important agreement
for void finders which accept the existence of galaxies in voids: The
overdensity of such galaxies, $\delta_{\mbox{\scriptsize{g20}}}$, is 
smaller than about $-0.8$, regardless of how voids are found.

There are also interesting agreements for quite different void finders.
For example, {\it Plionis/Basilakos} and {\it Pearce} find very similar results 
($FF_V$, $\delta_{\rm DM}$, and especially the position of the largest
void, but not $N_g$).

In Table~\ref{tab:results}, we also give the position of the centre of
the largest void and its radius, as provided by the different
groups. Note that some void finders build non--spherical voids, so the
quoted radius merely reflects the total size of the void.  While all
the finders indeed locate a large void within the central region it is
perhaps a little surprising that some centres are not within the
central structure that is so clearly visible in the top left
panel. This is in fact another consequence of the varying density
thresholds employed in that those finders with effectively lower
thresholds rely on larger scale structures than those that employ very
low density thresholds. In addition some methods (such as those of
{\it Brunino} and {\it Gottl\"ober}) find several voids of nearly equal size in
this region, as evidenced by the number of marked blue galaxies on
Figure~1 that are not within the marked green void. The key point is
that the filamentary structures visible in the top left panel of
Figure~1 are not very massive. Again it is clear that {\it void sizes
depend quite strongly on how voids are found, so one has to be very
careful about using void sizes to make statements about large--scale
structure.}

In Figures~\ref{fig:void_gals1} and \ref{fig:void_gals2}, we show void
galaxies found by the different groups. As noted above, the top left
and top centre panels of Figure~\ref{fig:void_gals1} give only the
dark matter distribution and the dark matter plus all model galaxies,
respectively.  All other panels superimpose all the recovered void
galaxies on top of the dark matter distribution. In addition, for each
group, we also show the largest void in green. The void centre is
marked with a large red dot.

As is clearly visible, 
there are quite large variations between the different groups, a
direct consequence of the wide variety of techniques and limits
employed. Hopefully these figures shed some light on the question of
what each group actually means when they refer to a ``void'' and
illustrate the inherent difficulty of comparing results obtained using
different void finders.
Individually the results of each group make perfect sense, when
seen in the light of how voids are identified. For example, the {\it Pearce}
voids are some of 
the most underdense spheres in the volume, centred on the particles
with the lowest density. Conversely, at first glance, the {\it Colberg} void doesn't
appear void at all and spans the entire figure, but this is a natural consequence
of the very low density of the entire region.

So unless agreement has been achieved on how to define what a void really is
-- or should be -- it is not straightforward to argue which void finder
does the best job, at least when comparing images. 

\subsection{Void Density profiles} \label{sec:profiles}

Despite the differences in the void--finding methods employed in earlier
studies, there has been broad agreement on two facts, namely that voids 
are very empty in their centres and that they have very sharp edges (see, 
for example, Benson et al. 2003, Colberg et al. 2005, or Patiri et al. 
2006b). Given the differences in the void finders, ``very empty'' might
mean different things. It might mean that the voids are literally empty
of the objects used as data -- such as galaxies below some given luminosity,
say -- or that voids do contain some material (for example dark matter), 
but very little of it.

With the large variety of void finders used here, it is an interesting and
important point to study the internal structure of a void. With the 
volume under consideration relatively small and underdense, most of the
void finders find one very large void, at about the same location. We are
thus limited to studying the structure of the largest void in each 
catalogue.

Figure~\ref{fig:dens_prof} shows the radially averaged enclosed dark
matter density as a function of radius for each of the catalogues,
using the void centres given in Table~\ref{tab:results} and shown
visually on Figures~1 and 2 as a red point. 
It 
is quite important to note that such a radial average is not ideal for void 
finders that produce non--spherical voids. Also, for each void finder 
the profile extends out to the largest radius that can be studied, given 
the size of the volume, so only the profiles of voids that lie close to the 
centre of the volume extend beyond 25\,$h^{-1}$\,Mpc. Note that these
radially averaged density profiles cannot be easily compared with the
average overdensities quoted in Table~\ref{tab:results}. The values
quoted in Table~\ref{tab:results} were computed using only the total 
void volume. However, radially averaging as in Figure~\ref{fig:dens_prof}
for voids that are not perfectly spherical will include material that
does not lie inside a void.

It appears that the void finders fall into three broad categories,
namely those which have central densities well below $\delta = -0.9$ 
({\it Foster/Nelson, Hahn/Porciani, Neyrinck, Platen/Weygaert, Plionis/Basilakos,
  Pearce}), those with much higher, flat, central densities
({\it Brunino, Gottl\"ober, Hoyle/Vogeley, M\"uller}) and a third set
  with intermediate central densities ({\it Colberg, Fairall, Shandarin/Feldman}).
Void finders with very low central densities all use the dark matter
density field in order to identify voids in combination with a low
effective overdensity threshold which restricts the size of the voids. 
The void finders that use
(model) galaxies or haloes all have somewhat higher central densities,
and much flatter central profiles. This effect is partly due to the
inclusion of small haloes near the void centres as well as the
difficulty of defining a void from a sparse sample of
points. Nevertheless it is clear that voids selected using the sparse
tracers available from galaxies or haloes typically have central
overdensities around $\delta = -0.85$ whereas those selected from the
richer dark matter distribution have typically lower central density
limits. Figure~3 further illustrates the role of the effective overdensity
threshold driving void choice: in the central region the method of
{\it Colberg} does not recover a particularly deep void, however, between
$15$ and $20h^{-1}$Mpc this method has found the most underdense
region of all the finders.

Up to a radius of around 15\,$h^{-1}$\,Mpc, the largest void in each
catalogue has an average density of $\delta \approx -0.85$ and at
larger radii the radially averaged densities are all rising.  However
the entire volume studied here has a mean $\delta = -0.28$ so none of
the voids runs into the very steep edges seen in earlier work as we
are still well below the mean cosmic density.

Despite the differences in the central densities, we can conclude that 
regardless of how voids are found, their interiors are very underdense
and they contain mean densities between 5\% and 20\% of the cosmic
mean. The central regions of voids also tend to have a rather flat
profile which means that regardless of how voids are found in
observational surveys, follow-up work of their interiors -- such as,
for example, searches for hydrogen (see, for example, Giovanelli et
al. 2005) -- should expect very low densities of material, provided,
of course, that the current model of structure formation used in 
the simulation is correct.

\subsection{Luminosity Function} \label{sec:lf}

Figure~\ref{fig:LF} shows the luminosity functions of galaxies in the 
entire Millennium simulation (solid red line), that of the
volume under consideration (solid black line) 
and in each of the catalogues of those void finders 
which identify galaxies inside voids, colour coded as shown on the figure. 

We present this plot mostly for illustrative purposes, since the
volume under consideration here is quite small. The key difference
between the full simulation and our selected subvolume is that the
galaxy formation efficiency across this entire region has been
suppressed.  In the full Millennium volume there are 7,151,282
galaxies with $M_B$ between -16 and -22. If the central
$40\,h^{-1}$\,Mpc of our subvolume was a random section of the full
box you would expect to find 3,661 galaxies. In practice our region has
707, or less than ${1 \over 5}$ of the expected number.  As well as
this overall normalisation, compared with the full volume of the
simulation, the luminosity function of the subvolume is very slightly
steeper at the faint end and is deficient in bright galaxies. As
mentioned before, the subvolume is underdense, so we do not expect to
find many bright galaxies.

The luminosity functions of the samples that contain significant numbers
of galaxies (with the exception of {\it Fairall}) show an even greater
deficiency of bright galaxies, as evidenced earlier by the
very low overdensity of bright galaxies in voids (Section~\ref{sec:numbers}).
Although it is difficult to tell, 
it looks as if at the fainter end, the luminosity functions of the void
samples all are just very slightly steeper than the subvolume one's and 
slightly steeper than the full simulation volume one's. This could be
seen as a trend towards the most isolated galaxies being fainter than
expected, as would be suggested from theoretical arguments. Those
galaxies residing in the most underdense regions (although there
aren't very many) are certainly faint. The limit of this effect, no
galaxies in the voids, is achieved by two finders, those of 
{\it Plionis/Basilakos} and {\it Shandarin/Feldman}. 

\subsection{Void galaxies and local environments} \label{sec:environments}

Given that we are interested in comparing results from different void
finders, it is worthwhile to look at which galaxies void finders identify
as belonging to a void. Apart from a galaxy's brightness, its environment,
expressed through some measure of the local density, provides a useful
descriptor. In order to quantify the local density, for each galaxy we 
compute $r_{14}$, the radius of the sphere that contains a mass of 
$10^{14}\,h^{-1}$\,M$_\odot$, roughly the mass of a small galaxy cluster.

In Figure~\ref{fig:local_dens}, we show the distribution of the values of
$r_{14}$, for both the complete subsample and the individual void galaxy
sets. Large (small) values of $r_{14}$ correspond to regions of low (high)
density. The distribution reflects the fact that the subvolume considered
here is underdense, since most galaxies reside in the low--density part
of the distribution. 

One would naively expect that void finders would pick up the galaxies 
in the lowest density regions first and then move towards the higher
density regions. However, while this is true for some of the void
finders, it is not true for all of them. This fact should be an important
criterion for future discussions of void finders: if a void finder 
locates galaxies inside voids, should these be those in the most
underdense environments?

Interestingly enough, the purely visual {\it Fairall} void--finding 
results in a distribution that is quite similar to {\it Colberg}'s, 
and also to {\it Neyrinck}'s and {\it M\"uller}'s. Given the
large differences in the methods these similarities are quite interesting,
and they merit to be taken into account in future discussions of
how to find voids.

\section{Summary and Discussion} \label{discussion}

This study represents the first systematic study of thirteen void finders, 
all of which have been used over the past decade to study voids, using the 
same data set to compare results. For the data we used real--space
coordinates of particles, haloes, and semi--analytical model galaxies 
(Croton et al. 2005) from a subvolume of the Millennium simulation 
(Springel et al. 2005). The goal of this paper was {\it not} to argue 
about the best way to define or identify voids. Instead, we aimed at 
allowing the reader to understand the differences between the methods to 
allow easier comparison of studies of voids in the literature. 

As outlined in Table~\ref{tab:voidfinders}, the void finders in this study
range from studies of the smoothed dark matter density field to identifying
empty spheres in the distribution of model galaxies, the latter either using
sophisticated algorithms or simply the human eye. Given the vastly different
assumption of what a void actually is, it is not surprising to see large
differences between some of the void finders. However, there are also some
quite encouraging agreements between methods that are quite different.

Not surprisingly, the different methods result in a large spread in basic
numbers such as the number of voids, the size of the largest voids (see
Table~\ref{tab:results}), or their basic appearance (see
Figures~\ref{fig:void_gals1} and \ref{fig:void_gals2}). 
We caution against putting too much emphasis on this fact. If one void 
finder constructs spherical voids with mean overdensities of $\delta = 
-0.9$ and another one builds large, irregularly shaped voids from spherical 
proto--voids in a distribution of galaxies, then the numbers and sizes of 
voids can be expected to be quite different. Likewise, the fraction of 
volume filled by the voids will be different. Regardless of these 
differences, it is quite interesting to see that the locations of the 
largest voids found by most of the groups agree quite well with each other. 
The eye finds a large void in the centre of the region studied, and the 
void finders do the same!

For a more detailed comparison the effective overdensity proves to be
most interesting.
Here, the spread is not quite as extreme as expected (see the values of
$\delta_{\mbox{\scriptsize{DM}}}$ in Table~\ref{tab:results}), and the
agreement in the overdensities of bright galaxies is quite impressive.
The void finders in this study agree that there should be no or just a
very small number of bright galaxies in voids. In other words, regardless
of how one defines voids, there are almost no bright galaxies in them.

As Section~\ref{sec:profiles} shows, the differences in the (radially 
averaged) density profiles of the largest void are also not very large,
with the void centres containing only between 5\% and 20\% of the 
mean density. This means that regardless of how voids are found, their
centres contain very little mass -- unless, of course, our model of
cosmic structure formation, which forms the basis of the simulation,
is wrong. With searches for HI emission in voids under way (see Giovanelli 
et al. 2005 or Basilakos et al. 2007), 
there should soon exist additional data points, which makes
it all the more pressing to move towards a more unified picture of voids.

As just mentioned, voids contain very few bright galaxies, and they contain
relatively more dim galaxies, something that those void finders that
identify void galaxies appear to agree on, too (see Section~\ref{sec:lf}).
Given the small number statistics in our sample, it is impossible to
make stronger comments about this. What appears clear, though, is that
this is an important topic to study, both observationally and theoretically,
in particular since current models of galaxy formation and evolution have
to account for the observed relation. For these studies to be successful,
more common ground is needed as far as defining and finding voids is
concerned.

We hope that this paper will trigger more detailed follow--up studies
to work towards a more unified view of how to define and find voids.
We believe that studies like this one, which make use of
high--resolution simulations of large--scale structure, provide
invaluable tools to this end, since they contain full information
about the distribution of model galaxies and of the underlying density
field. In the end, the model could then still be entirely wrong -- a
possibility that, in the light of the recent development of a standard
cosmological model, appears to be somewhat unlikely -- but it will
still be able to provide a sound basis for calibrations of methods and
ideas.  This point is of particular interest since observationally (at
present) only galaxies can be used to find voids. The distribution of
galaxies is much harder to model, though, than the cosmic density
field -- the latter can be described quite well using linear
theory. For studies of voids to be useful, a link needs to be forged
between theory and observation.  We hope that this work will provide a
basis for resolving this situation. Ultimately, as one of the most
extreme cosmic environments, voids possess the potential to constrain
models of galaxy formation. But for that to be the case, we need to
agree on what they really are and how to find them.

\section*{Acknowledgements}

This work was initiated at the Aspen Center for Physics' workshop on 
Cosmic Voids (June 2006) and at The Royal Netherlands Academy of Arts
and Sciences Colloquium on Cosmic Voids (December 2006). We thank both 
institutions for their generous and valuable support of this area of 
research. We also thank the Virgo Supercomputing Consortium and the
Lorentz Center in Leiden for their hospitality during the Winter
2007 Virgo Meeting.

The Millennium simulation used in this paper was carried out by the 
Virgo Supercomputing Consortium (http://www.virgo.dur.ac.uk) at the 
Computing Centre of the Max-Planck Society in Garching. We thank the Darren 
Croton for making the galaxy catalogue publicly available. Much of
the preparation and pre--analysis required for this work was carried
out at the Nottingham HPC facility.

J. Colberg is indebted to Carlos Frenk, Adrian Jenkins, and Lydia Heck for their 
support and help with Durham computing facilities, where part of this work 
was completed. C. Foster would like to thank NSERC (Canada) for support in 
the form of a graduate scholarship. L. Nelson would like to thank the 
Canada Research Chairs Program and NSERC for financial
support. Figure~1 was produced using Nick Gnedin's IFRIT software (http://home.fnal.gov/$\sim$gnedin/IFRIT).

\label{lastpage}

\end{document}